\documentclass[10pt]{article}

\usepackage[paper=a4,DIV=15]{typearea}
\usepackage[english]{babel}
\usepackage[utf8]{inputenc}
\usepackage{cite} 
\usepackage{authblk, amsmath,amsfonts,mathrsfs,amssymb,dsfont,braket,bbold,graphicx}
\usepackage{bm}
\usepackage[table,x11names]{xcolor}

\usepackage[export]{adjustbox}
\usepackage{tikz}
\usepackage[detect-all]{siunitx}
\usepackage{blindtext}
\usepackage{verbatim}
\usepackage{ellipsis}
\usepackage[autostyle]{csquotes}

\usepackage{enumitem}

\expandafter\let\csname equation*\endcsname\relax
\expandafter\let\csname endequation*\endcsname\relax

\usepackage{upgreek} 

\usepackage[unicode]{hyperref}

\usepackage[showonlyrefs]{mathtools}		
\mathtoolsset{showonlyrefs=true}

\newcommand{\e}[1]{\operatorname{e}^{#1}}

\newcommand{\I}{i}

\def\vec#1{\ensuremath{\mathchoice{\mbox{\boldmath$\displaystyle#1$}}
                            {\mbox{\boldmath$\textstyle#1$}}
                            {\mbox{\boldmath$\scriptstyle#1$}}
                            {\mbox{\boldmath$\scriptscriptstyle#1$}}}}

\def\ba#1\ea{\begin{equation}\begin{aligned}#1\end{aligned}\end{equation}}																

\usepackage{booktabs}
\AtBeginDocument{
\heavyrulewidth=.08em
\lightrulewidth=.05em
\cmidrulewidth=.03em
\belowrulesep=.65ex
\belowbottomsep=0pt
\aboverulesep=.4ex
\abovetopsep=0pt
\cmidrulesep=\doublerulesep
\cmidrulekern=.5em
\defaultaddspace=.5em
}
\usepackage{threeparttable}
\usepackage{colortbl}
\definecolor{Gray}{gray}{0.85}

\begin{document}

\title{Realizing Quantum free-electron lasers: A critical analysis of experimental challenges and theoretical limits}

\author[1,*]{Alexander~Debus}
\author[1]{Klaus~Steiniger}
\author[2,1]{Peter~Kling}
\author[1,2]{Moritz~Carmesin}
\author[1,3]{Roland~Sauerbrey}

\affil[1]{Helmholtz-Zentrum Dresden -- Rossendorf e.V. (HZDR), Bautzner Landstraße 400, D-01328 Dresden, Germany}
\affil[2]{Institut für Quantenphysik and Center for Integrated Quantum Science and Technology, Universität Ulm, Albert-Einstein-Allee 11, D-8981, Germany}
\affil[3]{Technische Universität Dresden, 01062 Dresden, Germany}
\affil[*]{E-mail: \href{mailto:a.debus@hzdr.de}{a.debus@hzdr.de}}

\date{\today}

\maketitle

\begin{abstract}
We examine the experimental requirements for realizing a high-gain Quantum free-electron laser (Quantum FEL). Beyond fundamental constraints on electron beam and undulator, we discuss optimized interaction geometries, include coherence properties along with the impact of diffraction, space-charge and spontaneous emission.
Based on desired Quantum FEL properties, as well as current experimental capabilities, we provide a procedure for determining a corresponding set of experimental parameters.
Even for an idealized situation, the combined constraints on space-charge and spontaneous emission put strong limits on sustaining the quantum regime over several gain lengths. Guided by these results we propose to shift the focus towards seeded Quantum FELs instead of continuing to aim for self-amplified spontaneous emission (SASE). Moreover, we point out the necessity of a rigorous quantum theory for spontaneous emission as well as for space-charge in order to identify possible loopholes in our line of argument.
\end{abstract}

Recent advances regarding x-ray free-electron lasers in the \AA-wavelength range \cite{Emma2010,Ishikawa2012,Reschke2017,Milne2017} have led to tremendous progress in revealing the structure of matter in physics, chemistry, material science and the life sciences \cite{Seddon2017}.
However, current x-ray FELs are large, expensive, km-long facilities that can offer access to only a few selected experiments.

The dream of a high-gain, Quantum SASE FEL combines compact devices on the scale of a university laboratory with unprecedented small FEL bandwidths down to the transform limit at both full transverse and longitudinal coherence.
Especially the latter feature is not yet available in existing machines \cite{Yabashi2017,Pellegrini2016,Barletta2010,McNeil2010,Kim2008} and would be a catalyst for opening another window into the structure of matter.
Compactness of Quantum FELs arises from the use of laser pulses as optical undulators~\cite{sukhatme,pantell,schlicher,schlicher_ieee,sprangle2009,Steiniger2014_3}.
These reduce both the undulator length to centimeters and the electron energy $E_\mathrm{e}$ to a few ten megaelectronvolt for the production of coherent x-ray pulses due to their micrometer scale undulatorperiods.

These prospects have put the Quantum FEL (QFEL) into the focus of continuous theoretical interest \cite{Bonifacio2005a,boni06,boni_wigner,bonifacio-basis,eli2,Robb2011,Monteiro2011,Robb2013,Kling2015,applb,Kling2017,Brown2017,anisimov2018,gover2018dimension} since 2005.
These studies showed that the quantum regime of the FEL emerges when the discrete recoil $\hbar k_\text{FEL}$, which an electron experiences when it scatters from a photon, exceeds
the classical gain bandwidth $\rho E_\mathrm{e}$.
However, studies concerned with the actual challenges towards experimental realization have been scarce \cite{Bonifacio2005b,BONIFACIO2007,Bonifacio2017}.
Often these analyses point to the technical challenges of providing the required electron beam quality or drive laser amplitude, spot size and pulse length.
Beyond selected parameter sets, critical problems of Quantum FELs arising during interaction, such as transverse coherence, space-charge or spontaneous emission were investigated in isolation without estimating scalings and systemic interdependencies.

In this article we provide an overview on the basic QFEL theory, as well as a list of requirements and scaling laws in order to identify suitable regimes for experimental realization.
We discuss in detail the benefits and implications of using an optical laser as undulator for a QFEL.
In addition to the challenges of providing electron bunches with ultra-low energy spread and low normalized emittance, we describe two critical limitations for Quantum SASE FEL:
First, we show that requirements on intensity and phase stability of the optical undulator field are unnecessarily restrictive for Quantum FELs if a standard focal geometry of a Gaussian laser pulse colliding head-on with an electron beam is used.
We propose that this challenge can be met by utilizing Traveling-Wave Thomson-Scattering (TWTS) which makes use of a side-scattering geometry and pulse-front tilted laser pulses.

Secondly, we derive that even for an idealized situation the combination of spontaneous emission and space-charge constraints imposes strong limits on sustaining the quantum regime over several gain lengths.
This is our central result which appears to be prohibitive for all Quantum SASE FELs aiming at the generation of fully coherent x-ray beams from electron beam shot-noise until saturation.
Best-case estimates motivated by classical FEL theory including space-charge show that the deep quantum regime $\bar\rho \ll 1$ remains out-of-reach, where $\bar\rho$ denotes the quantum parameter of ref.\ \cite{boni06}.
Only regimes with a quantum parameter $\bar{\rho}\geq 0.5$ would be directly accessible using a Quantum SASE FEL.

Therefore, we close the article by proposing a shift of theoretical and experimental efforts towards seeded Quantum FELs instead of self-amplified spontaneous emission.
This step can lead to novel applications with respect to FEL amplifiers in the hard x-ray range and diagnostics for electron of extreme 6D-brightness and highly coherent x-ray beams. 

Aiming for a deeper understanding, as well as for testing our argumentation against potential loopholes, we stress the need for a complete quantum theory of the Quantum FEL, which includes the effects of both spontaneous emission and space-charge.

\section{Introduction: What is a QFEL?}

A Quantum FEL denotes a special regime of free-electron laser operation.
Like its classical counterpart, it consists of a relativistic electron beam with an undulator field from either a magnetic or optical undulator, as well as a resulting FEL field.
The electrons interact with both the undulator field and FEL radiation.
Provided that electron beam and undulator field fulfill some threshold criteria with respect to beam and field quality a collective instability emerges, in which a part of the kinetic energy $E_\mathrm{kin}$ of the electrons is converted to FEL radiation.
Due to the relativistic nature of the interaction, the emitted light is Doppler-upshifted up to the hard x-ray range.

In contrast to a classical FEL, where the electrons travel on deterministic trajectories, the electron motion in a Quantum FEL is characterized by discrete momentum steps which occur only with a certain probability according to Born's rule.
These discrete jumps emerges when an electron scatters an undulator and an FEL photon whereby it recoils by about $\hbar k_\mathrm{FEL}$.
If this recoil is small compared to the absolute momentum bandwidth of the classical FEL, the discreteness of the dynamics is washed out. 
This case corresponds to the classical regime.

The quantum regime~\cite{banacloche,friedman,becker88,kurizki} of the FEL will be entered if the recoil $\hbar k_\mathrm{FEL}$ is large compared to the absolute momentum bandwidth of the classical FEL $\rho \gamma m c$.
This condition leads to the definition of the dimensionless QFEL parameter~\cite{boni06}
\begin{equation}
  \bar{\rho} \equiv \rho \frac{\gamma m c}{\hbar k_\text{FEL}} = \rho\gamma \frac{\lambda_\text{FEL}}{\lambda_c}\, ,
  \label{eq::QFELparameter}
\end{equation}
where $\gamma = E_\mathrm{kin}/mc^2$, $m$, $k_\text{FEL}$, $\lambda_c\equiv h/(mc)$, $\lambda_\text{FEL}$, $\rho$ denote the dimensionless electron energy, electron mass, FEL wavenumber, Compton wavelength, FEL wavelength and the dimensionless FEL parameter respectively.
We emphasize that the quantum regime can be accessed only if $\bar{\rho}\ll 1$.

\begin{figure}[!t]
  \centering
  \begin{tikzpicture}
    \node[anchor=north west,inner sep=0] at (0,0) {\includegraphics[height=6.0cm]{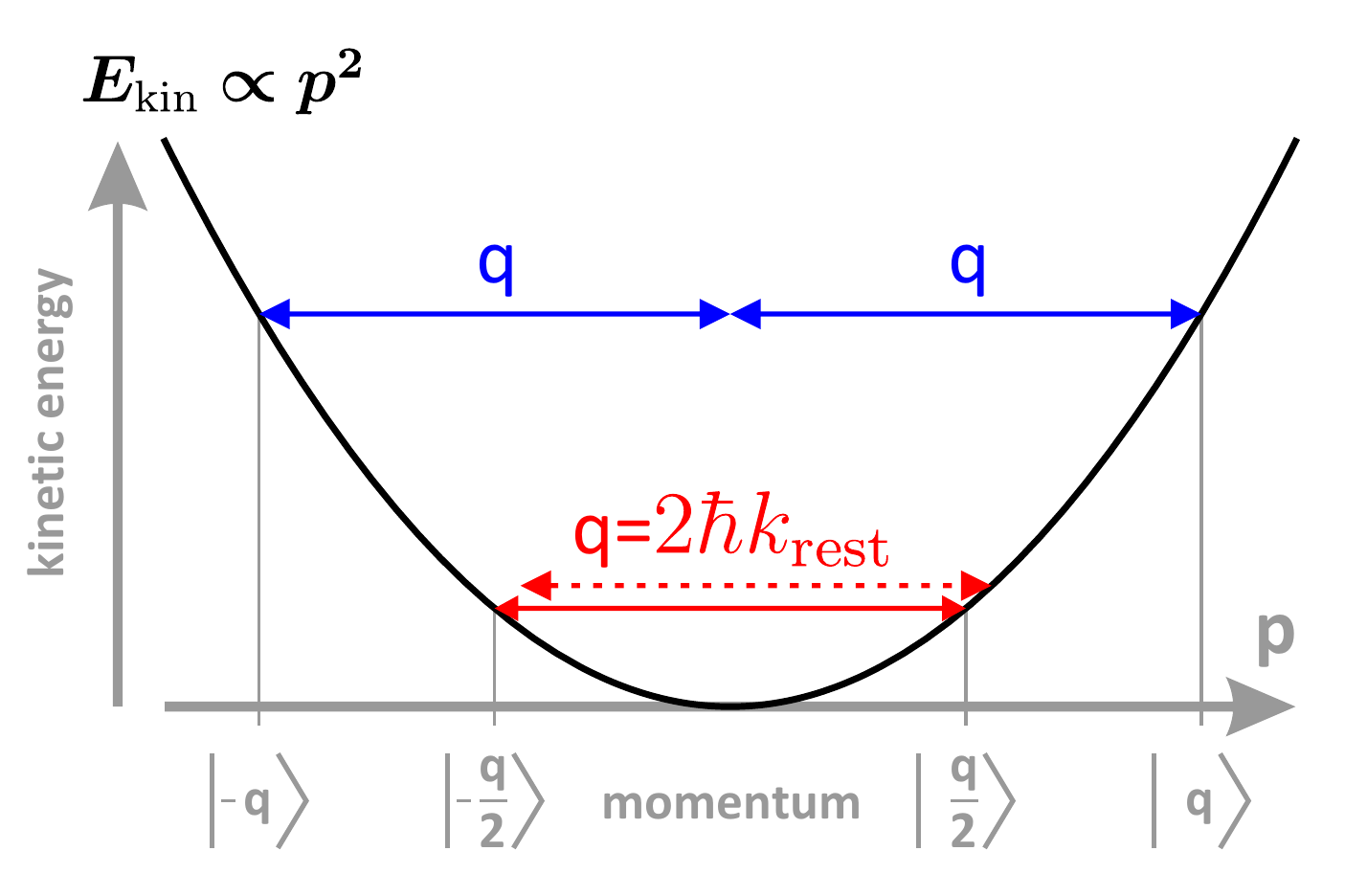}};
    \node[right,align=left] at (0.2,0.0) {(a)};
  \end{tikzpicture}
  \\
  \begin{tikzpicture}
    \node[anchor=north west,inner sep=0] at (0,0) {\includegraphics[height=6.0cm]{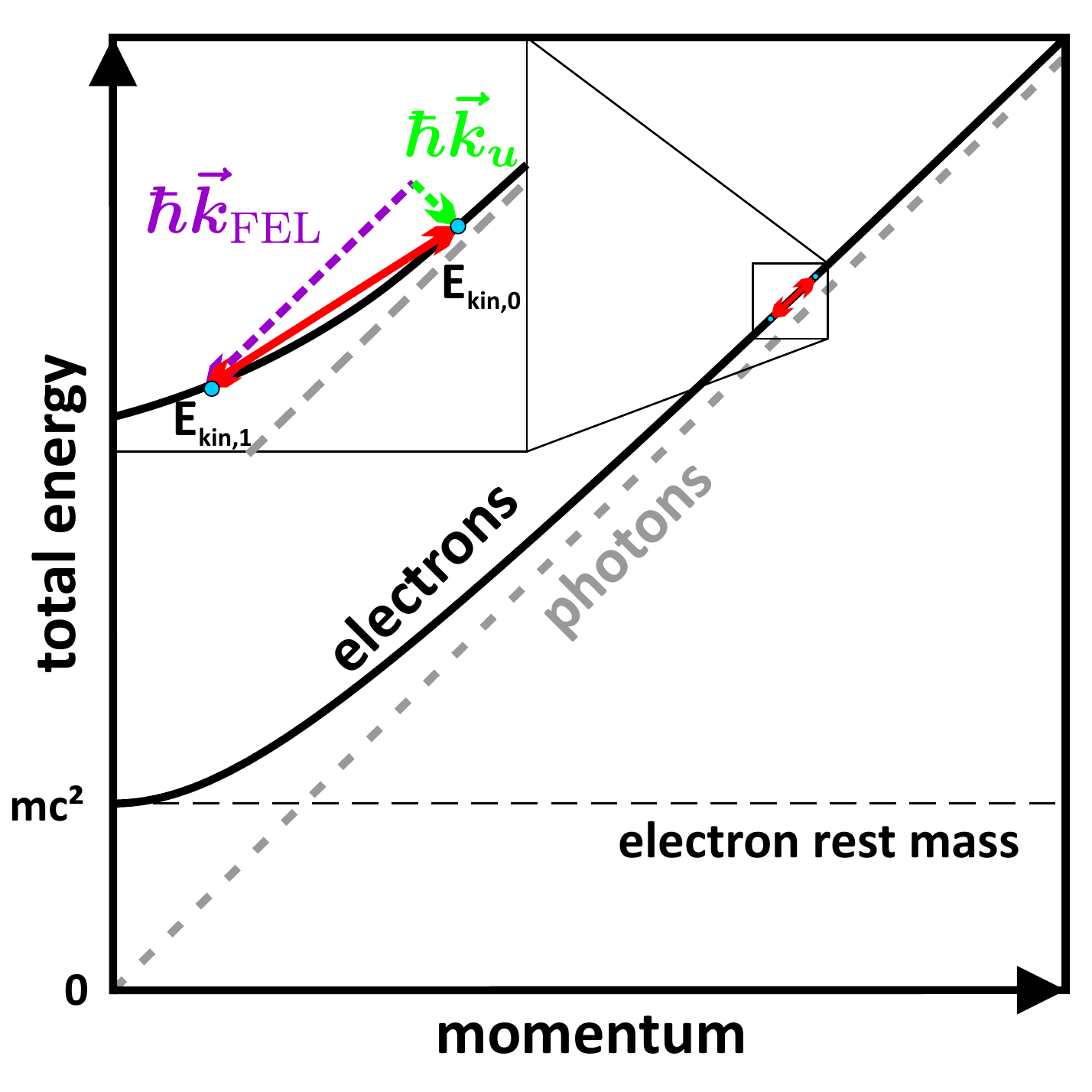}};
    \node[right,align=left] at (-0.2,0.0) {(b)};
  \end{tikzpicture}
  \hspace{1cm}
  \begin{tikzpicture}
    \node[anchor=north west,inner sep=0] at (0,0) {\includegraphics[height=6.0cm]{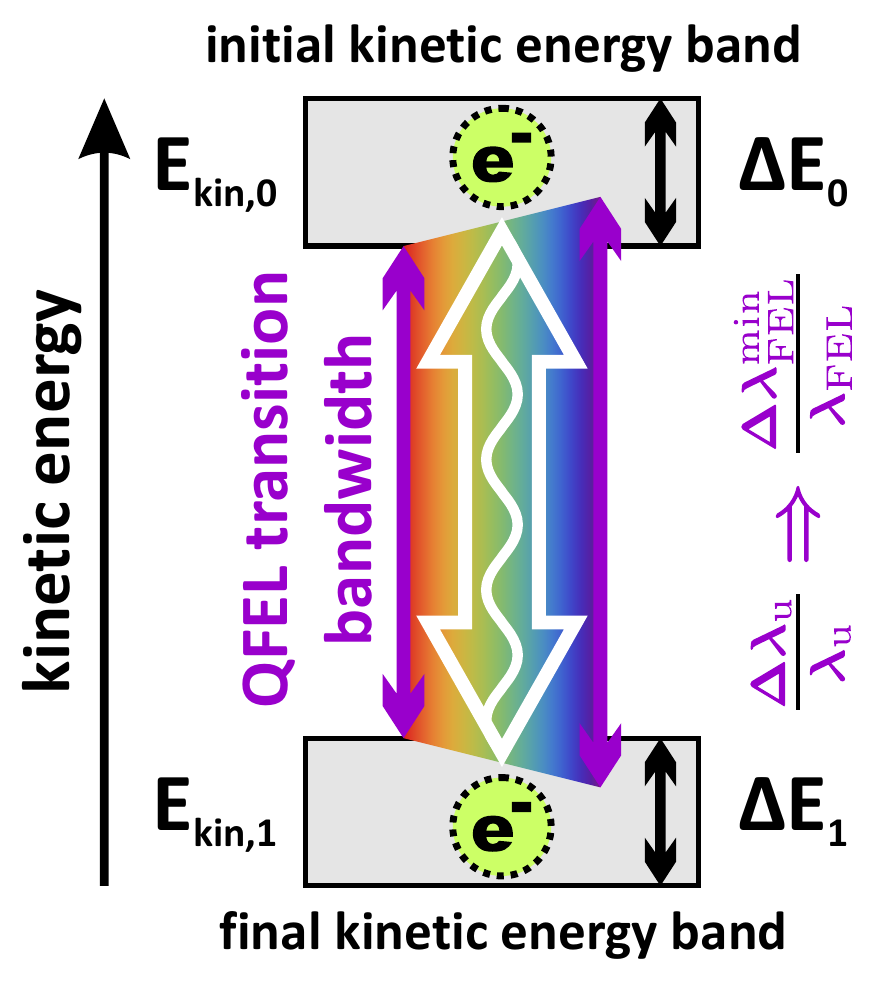}};
    \node[right,align=left] at (-0.2,0.0) {(c)};
  \end{tikzpicture}
  \caption{(a) depicts electron energy and momentum with emission and absorption events of undulator and FEL photons in the average rest-frame (Bambini-Renieri frame), where $k_\text{rest}\equiv k_u\equiv k_\text{FEL}$.
  At small energy and momentum uncertainties, as well as small laser bandwidths compared to the photon recoil, only selected transitions are allowed.
  The fundamental transition (red, solid) is more probable than higher-order transitions (blue, solid) or off-resonant transitions (red, dashed).
  (b) shows the corresponding situation for a relativistic electron in the laboratory frame, where $k_u\neq k_\text{FEL}$ and the electron energy-momentum relation reads $\gamma(p)=\sqrt{1+(p/mc)^2}$.
  (c) emphasizes the finite bandwidths of the electron momentum states and the laser transition bandwidths.
  Opposed to two-level systems such as in atomic bound states, both initial momentum bandwidth and undulator bandwidth are not \emph{a priori} defined and have to be externally enforced by the experimental setup for the duration of the entire interaction.
  In addition, every real-world (laser) undulator features a finite bandwidth and thus defines the minimum relative bandwidth of the resulting QFEL radiation.}
  \label{fig::QFELIntro}
\end{figure}
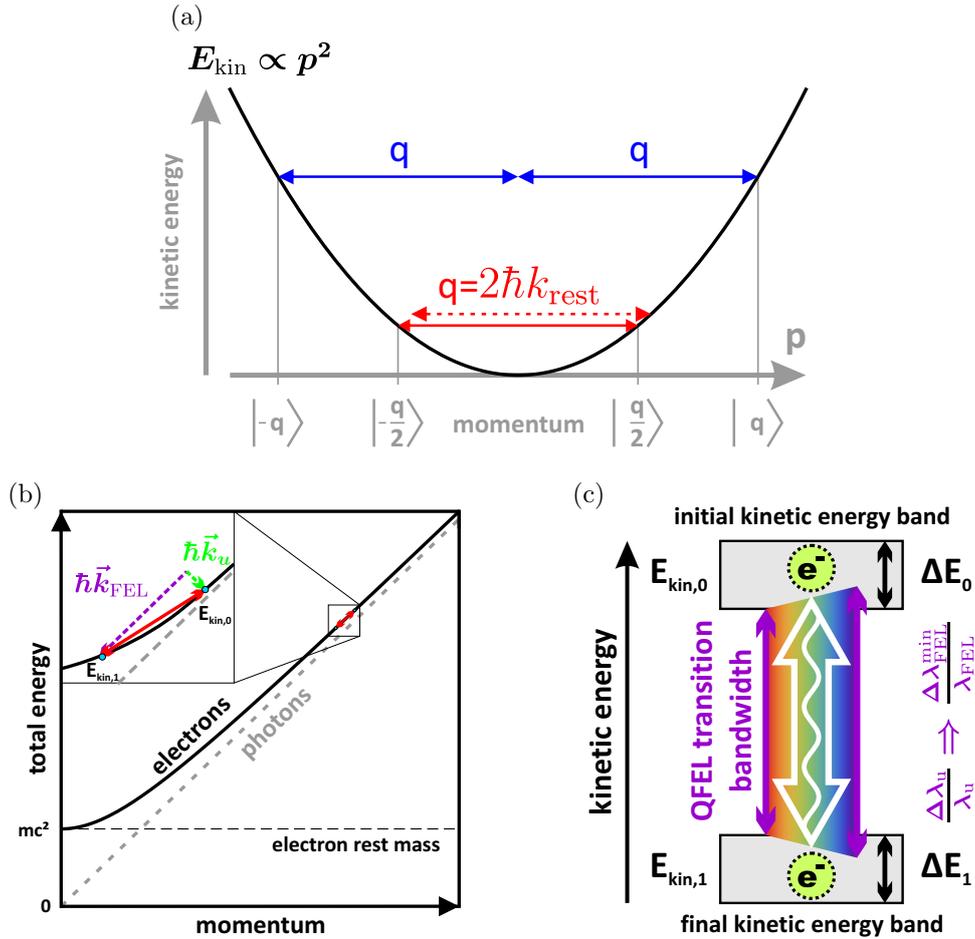

This quantum mechanical recoil can now be exploited to construct and isolate a two-level quantum system.
This becomes apparent in the average rest frame of an electron, Fig.~\ref{fig::QFELIntro}(a), where the FEL and the undulator field have equal frequency $\omega_\text{rest}$.
The energy-momentum relation of an electron at small velocities is a parabola, while the dispersion relation of the fields is linear.
If momentum uncertainties of electron and photon are negligibly small, only discrete transitions are allowed according to Fig.\ref{fig::QFELIntro}(a): One undulator photon is absorbed by the electron, which in turn gains the discrete energy and momentum of the photon.
At the same time an FEL photon is emitted in opposite direction reducing the electrons energy and momentum by the discrete photon energy and photon momentum, $\hbar \omega_\text{rest}$ and $\hbar k_\text{rest}$ respectively.
If in addition the intensity of the FEL radiation is small, multiphoton transitions are much less likely to occur than processes where a single photon is emitted or absorbed.
Hence, a single, discrete transition becomes the dominant mode of interaction, which constructs a system of two momentum levels~\cite{Kling2015}.
Exploiting the quantum mechanical recoil together with energy and momentum conservation therefore excludes transitions to other momentum levels.
Especially it is not possible within this two-level system to extract more energy than $\hbar k_\text{FEL} c$ per electron.

The interaction can also be described in the laboratory frame~\cite{applb}, as shown in Fig.\ref{fig::QFELIntro}(b).
An electron with initial energy $\gamma$ gains energy by absorbing an undulator photon, while loosing the momentum $\hbar k_u$ due to the opposite direction of light propagation.
At the same time, the electron looses energy by emitting an x-ray FEL photon in the direction of electron propagation causing again a loss of momentum, but now by a value of $\hbar k_\text{FEL} \gg \hbar k_\mathrm{u}$.
In total this two-photon process defined by the FEL resonance condition $k_\text{FEL}\simeq 2\gamma^2 k_u$ uniquely connects two energies by a secant.
Assuming that higher harmonics processes are negligible, quantum mechanics thus enforces a selection rule for the fundamental two-photon process that specifically forbids energy gain to energies above $E_{\mathrm{kin},0}$, as well as energy losses to energies below $E_{\mathrm{kin},1}$.

Due to a finite bandwidth of the undulator laser, quantum uncertainties of a single electron or the classical momentum spread of the electron bunch, the two points of intersection of the secant are in reality small, but finite energy and momentum bands.
However, if either these finite bands or the transition bandwidth determined by the undulator field become large enough to enable several secants on the energy scale or even merge all these bands to one single band, quantum effects are washed out leading to the physical make-up of a classical FEL.
Qualitatively speaking, the quantum regime of an FEL can only be accessed through an experimental setup, where the electron distribution as well as the spectrum of the optical undulator allow only the fundamental two-photon transition between two discrete and disjunct sets of energy and momentum bands, see Fig.~\ref{fig::QFELIntro}(c).

\begin{figure}[!t]
  \centering
  \begin{tikzpicture}
    \node[anchor=north west,inner sep=0] at (0,0) {\includegraphics[height=7.5cm]{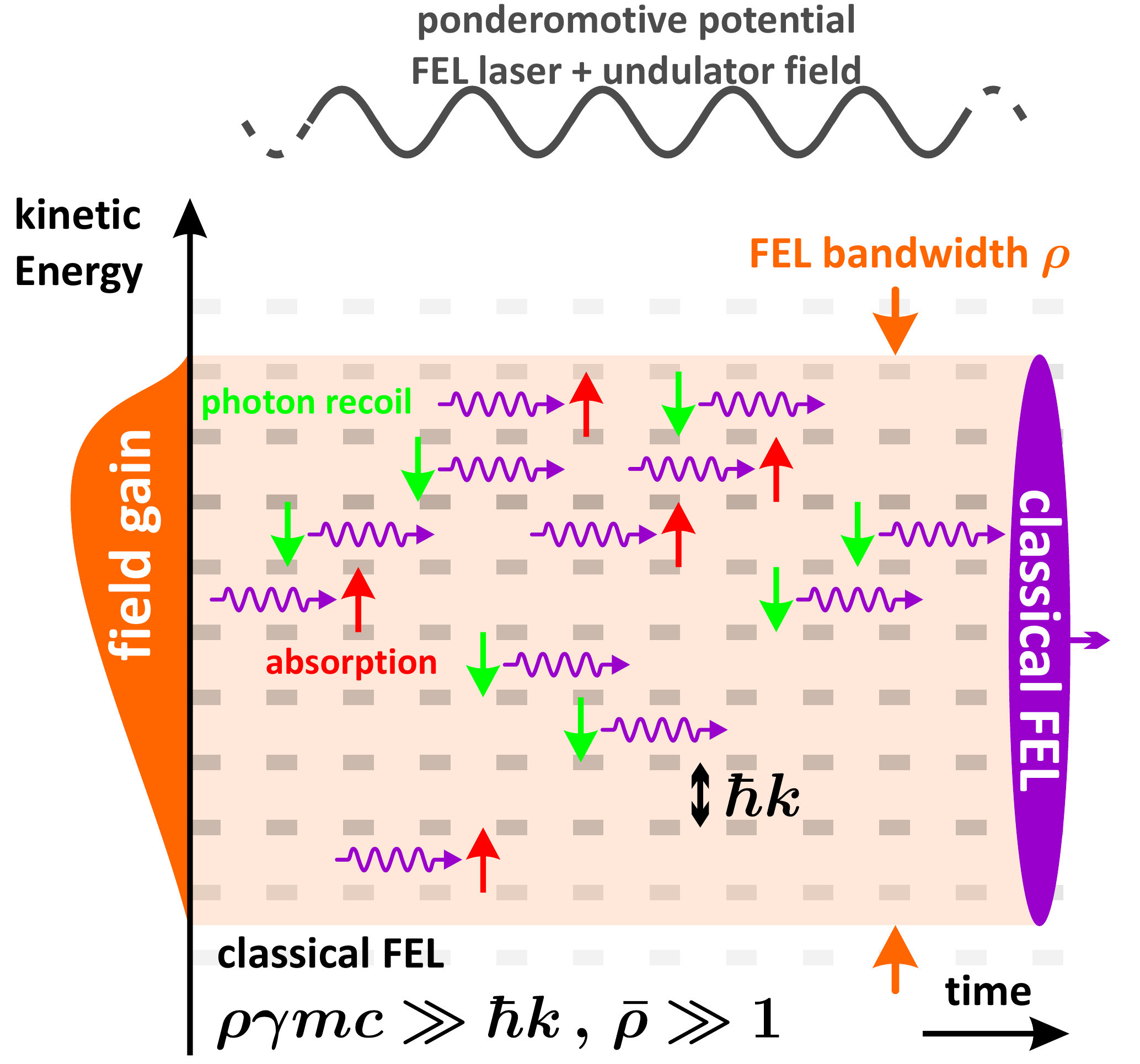}};
    \node[right,align=left] at (0.0,0.0) {(a)};
  \end{tikzpicture}
  \hfill
  \begin{tikzpicture}
    \node[anchor=north west,inner sep=0] at (0,0) {\includegraphics[height=7.5cm]{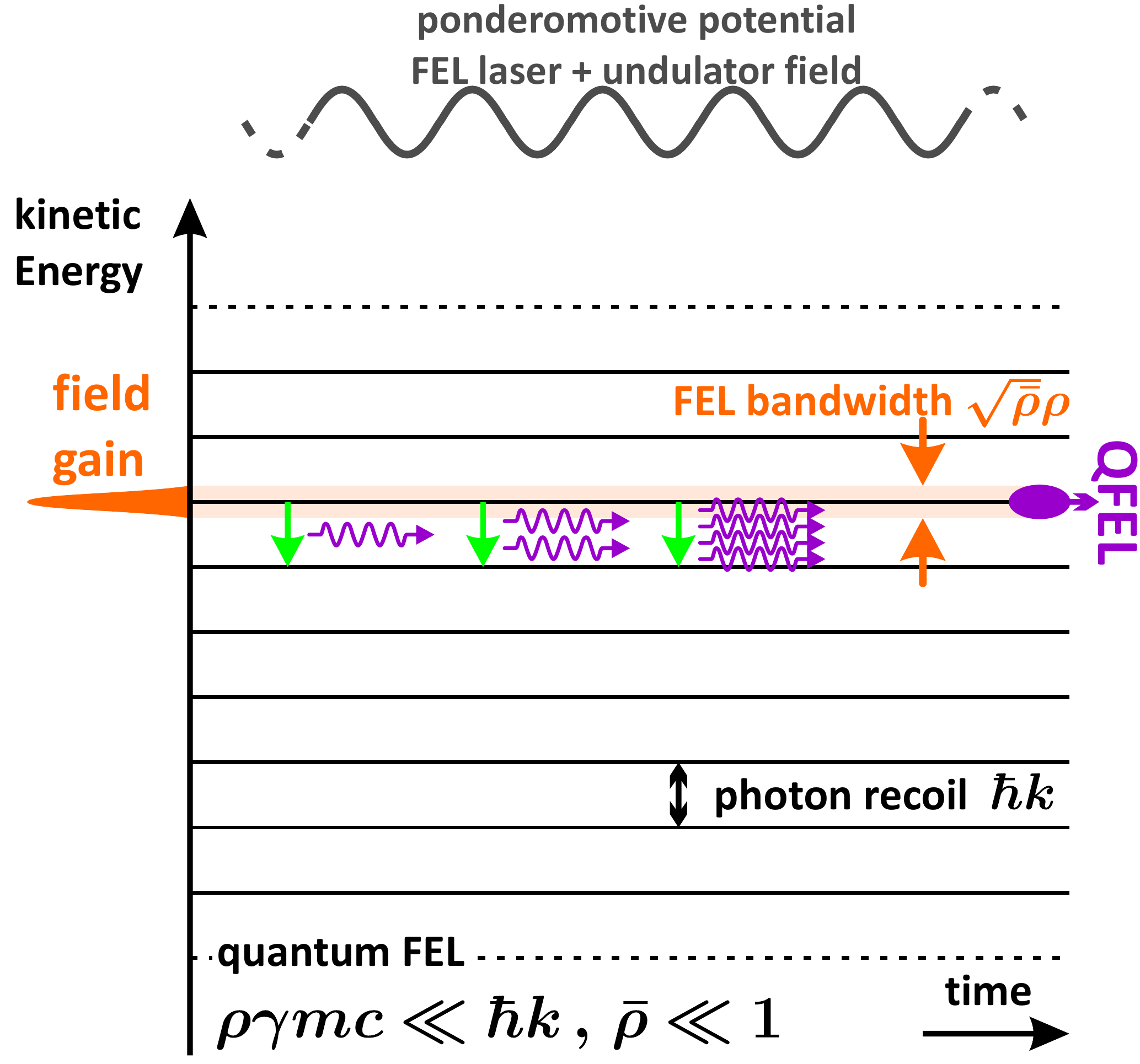}};
    \node[right,align=left] at (0.0,0.0) {(b)};
  \end{tikzpicture}
  \caption{(a) depicts the evolution of a classical FEL, where the quantum mechanical recoil is much smaller than the relative FEL bandwidth $\rho$.
  Potentially discrete momentum levels overlap, so that phase space is continuous.
  All electrons can have multiple photon emission and absorption events during the FEL interaction.
  (b) shows the evolution of a Quantum FEL where the quantum mechanical recoil is larger than the relative Quantum FEL bandwidth $\sqrt{\bar{\rho}}\rho$, with $\bar{\rho}\ll 1$.
  In this limit momentum bands are partitioned and isolated in phase space as well as a single transition selected.
  This constructs a two-level system.
  By a suitable choice of electron energy and undulator wavelength electrons start in the upper energy level. FEL lasing happens through stimulated emission in the interaction with the undulator and x-ray fields. Note that energy bands are only approximately equidistant. According to $E_\text{tot}=\gamma mc^2$ and Fig.~\ref{fig::QFELIntro}(b) it is actually critical that such a weak energy dependency exists.}
  \label{fig::ComparisonQFELClassicalFELDynamics}
\end{figure}

A classical FEL as depicted in Fig.~\ref{fig::ComparisonQFELClassicalFELDynamics} has either multiple energy bands due to $\bar{\rho}\gg 1$ or no discrete band structure at all due to quantum mechanical uncertainties, finite bandwidth of the electron energy distribution as well as the undulator field, or from competing physics processes such as space-charge and spontaneous emission.
In all these cases, each electron can absorb and emit several photons~\cite{fedorov_rev} within the energy bandwidth $\rho\gamma mc^2$ during the classical FEL interaction.
Over the statistics of many interactions, the mean energy loss of the electron ensemble translates to FEL photon yield and can be several photons per electron.

A Quantum FEL $\bar{\rho}\ll 1$, corresponding to a large quantum mechanical recoil compared to the classical FEL bandwidth $\rho$, allows only a single class of transitions between two energy bands as shown in Fig.~\ref{fig::ComparisonQFELClassicalFELDynamics}(b).
By initial choice of electron energy and undulator wavelength, electrons start in the upper energy band and exhibit Rabi oscillations~\cite{schleich} between the two energy bands.
At best, if the QFEL interaction is ended when all electrons reside in the lower energy level, each electron has emitted a single FEL photon.
Although each electron can only emit one single FEL photon, the photon statistics evolves continuously according to the interacting electrons and photons.
In the same way as for optical solid-state lasers, the stimulated emission leads to a self-reinforcing avalanche until the upper energy level is depleted and all electrons reside in the lower energy band.
Considering a two-level system, it is necessary to end the interaction at the right moment, otherwise the electrons will continue to oscillate between the two energy levels until either the FEL radiation has spatially left the interaction region by slippage or non-ideal effects end the interaction.

For the sake of completeness, we note that according to Refs.~\cite{boni06,anisimov2018} a micro-bunching of electrons also exists in a QFEL.
Unlike the classical FEL, where the micro-bunching emerges by the motion of the electrons in a ponderomotive potential, the corresponding effect in the quantum regime can be understood by means of position and momentum uncertainty~\cite{anisimov2018}.

\section{Theory of the Quantum FEL}

In the following, we motivate the most important conditions and features of a Quantum FEL. The starting point of this discussion is the intuitive model in Ref.~\cite{Kling2015}, where the emergence of the quantum regime of the FEL was studied from a quantum optics point of view. Although this model is based on a low-gain theory, it enables us to identify the most important quantities for the Quantum FEL even in the high-gain regime.
We note that a rigorous high-gain theory is out of the scope of the present article, but will be presented elsewhere.

Instead, we highlight the connection of the different quantities between the formalisms established in Ref.~\cite{Kling2015} and Ref.~\cite{boni06}.
While the former approach is focused on the fundamental aspects, models based on the latter one tend to employ concepts of classical FEL theory and are therefore closer to the usual notation of standard FEL literature.
For this purpose, we connect to the usual notation and scaling of parameters~\cite{boni06}.

\subsection{Basic quantum model}

The dynamics of electron, radiation field and undulator field in the average rest frame of the electron, see Fig.~\ref{fig::QFELIntro}(a), the so-called Bambini-Renieri frame~\cite{bambi,brs}, is dictated by the Hamiltonian~\cite{becker82,becker83}
\ba\label{eq:qtheory_H}
\hat{H}=\frac{\hat{p}^2}{2m}+\hbar \tilde{g} \left(\hat{a}_\text{FEL}\hat{a}_\text{u}^\dagger\e{\I2k_\text{rest}\hat{z}}+\hat{a}_\text{FEL}^\dagger\hat{a}_\text{u}\e{-\I2k_\text{rest}\hat{z}}\right)\,,
\ea
where $\hat{z}$ and $\hat{p}$ denote the position operator of the electron along the wiggler axis and the conjugate momentum operator, respectively, while $\hat{a}_\text{FEL}$, $\hat{a}_\text{FEL}^\dagger$ and $\hat{a}_\text{u}$, $\hat{a}_\text{u}^\dagger$ are the photon annihilation and creation operators for the radiation and the undulator field, respectively. The constant $\tilde{g}$ describes the coupling of electron and fields and incorporates the vacuum amplitudes of the fields as well as the mass $m$ of the electron.

Although $\hat{H}$ has an analogue form as the classical pendulum Hamiltonian we emphasize that its consequences are fundamental different from a classical theory: while the classical Hamiltonian equations are solved by deterministic trajectories, solutions to the Schrödinger equation subject to the Hamiltonian $\hat{H}$, Eq.~\eqref{eq:qtheory_H}, are characterized by \textit{probability amplitudes}, implying that transitions between momentum levels occur only with a certain probability.

By considering the action of the single contributions of $\hat{H}$, Eq.~\eqref{eq:qtheory_H}, on a quantum state we again observe momentum conservation like in Fig.~\ref{fig::QFELIntro} (a), but now on a more formal level. For example, we obtain
\ba
\hat{a}_\text{FEL}^\dagger\hat{a}_\text{u}\e{-\I 2k_\text{rest}\hat{z}}\ket{n_\text{FEL},n_\text{u},p}=
\ket{n_\text{FEL}+1,n_\text{u}-1,p-2\hbar k_\text{rest}}\,.
\ea
While one photon is emitted into the radiation field and one is absorbed from the undulator field,
the momentum of the electron is decreased by the recoil $q\equiv 2\hbar k_\text{rest}$. Here we have assumed a product state consisting of two Fock states with photon numbers $n_\text{FEL}$ and $n_ \text{u}$ for the radiation and undulator field, respectively, and of one momentum eigenstate for the electron with the momentum $p$. In general we have to deal with a superposition~\cite{ciocci} of all possible transitions yielding
\ba
\ket{\Psi}=\sum\limits_\mu c_\mu \ket{n_\text{FEL}+\mu,n_\text{u}-\mu,p-\mu q}\,,
\ea
where $c_\mu$ denotes the probability amplitude for finding the system in one of these configurations which can be distinguished by their quantum number $\mu$ which simultaneously indicates the number of scattered photons into or from the respective field and the number of discrete momentum jumps of the electron.

Before we proceed, we emphasize that the model leading to the Hamiltonian in Eq.~\eqref{eq:qtheory_H} is quite basic: it covers only one electron, one spatial dimension and two modes of the radiation field. In order to model realistic experiments, one would need to consider a continuous distribution of both Fock states and electron momentum states as well as three-dimensional effects. However, our simple approach still proves to be a powerful tool to derive and understand important conditions and parameters for a Quantum FEL as we show in the following.

\subsection{Quantum parameter}

The crucial step in deriving the fundamental Quantum FEL dynamics \cite{Kling2015} is to identify the two important frequency scales of the FEL dynamics, which are
(i) the coupling strength $g\sqrt{n_\mathrm{FEL}}$ and (ii) the recoil frequency
\ba
\omega_\text{rec}\equiv \frac{1}{\hbar}\frac{q^2}{2m}\,,
\ea
which is the energy associated to one momentum step with $q\equiv2\hbar k$ divided by $\hbar$.
We note that we have redefined the coupling $\tilde{g}$ as $g\equiv \tilde{g}\sqrt{n_\text{u}}\propto a_0$ which is proportional to the undulator parameter $a_0$, eq.~\eqref{eq::a0HandyDef}. This rescaling was possible since the number $n_\mathrm{u}$ of photons in the undulator field is very large and we can approximate it by a constant. Thus, we are allowed to remove the degree of freedom associated to the undulator from the dynamics~\cite{becker82,becker83}.

The quantum regime is entered, if the ratio of these two frequencies, defined as the quantum parameter
\ba
\alpha\equiv \frac{g\sqrt{n_\mathrm{FEL}}}{\omega_\text{rec}}
\ea
is small, that is $\alpha \ll 1$.
We can understand this statement in the following way: in the equations of motions~\cite{Kling2015}, the transitions between different momentum levels are characterized by  oscillations with different multiples of the recoil frequency $\omega_\text{rec}$. For $\alpha\ll 1$ the recoil frequency is large and oscillations with multiples of it can be considered as rapid ones.
Hence, we neglect these fast varying terms in a rotating-wave like approximation~\cite{schleich}. Only the dynamics of the transition from
$p=q/2$ to $p=-q/2$ is independent of $\omega_\text{r}$ and thus is slowly varying.
Hence, only this transition survives and thus  forms the basis of the two-level approximation for the quantum regime~\cite{NJP2015}.

We note that by employing asymptotic methods~\cite{bogoliubov} we observe that the probability of transitions outside the two-level system of $q/2$ and $-q/2$, that is, processes, where more than one photon is scattered into the FEL mode, scale with integer powers of the quantum parameter $\alpha$ in comparison to single-photon processes and thus are suppressed in the quantum regime, $\alpha \ll 1$.
Hence, employing the parameter $\alpha$ is the natural choice for understanding the Quantum FEL from first principles.

However, for experimental considerations it is more convenient to use the parameter
\ba\label{eq:qtheory_rhobardef}
\bar{\rho}\equiv \rho \frac{mc\gamma}{\hbar k_\text{FEL}}
\ea
from Ref.~\cite{boni06} which includes the FEL parameter $\rho$ and is written in terms of the laboratory frame. In Ref.~\cite{Kling2015} a one-to-one relation between the two different representations was established, which reads~\footnote{We note that in Eq.~(40) of Ref.~\cite{Kling2015} an additional prefactor of $\sqrt{2}$ appears. This discrepancy has emerged since in this reference a slightly different definition of the FEL parameter $\rho$ was used.}
\ba\label{eq:qtheory_one_to_one}
\alpha\equiv \bar{\rho}^{3/2}\,.
\ea
In order to establish this connection of $\alpha$ and $\bar{\rho}$ we had to set the number $n_\mathrm{FEL}$ of photons equal to the number $N$ of electrons.
This estimate is motivated by the fact that in the quantum regime each electron emits maximally only one photon.
We note that this rather hand-waving procedure is inherent to the low-gain approach of Ref.~\cite{Kling2015} and would become unnecessary in a proper high-gain theory were $\alpha$ should emerge in a description of the collective electron dynamics.

In contrast to $\alpha$, which appears only in integer powers in the equations of the quantum regime, the $3/2$-scaling of $\bar{\rho}$ in eq.~\eqref{eq:qtheory_one_to_one} has its source in geometry and dimensional considerations of FEL physics, especially from transverse and longitudinal dynamics.
It originates from the description of the collective interaction of many electrons with the fields and quantifies the strength of the electron-light coupling.
As it represents the growth rate of the microbunching instability, it can be seen as a measure of the efficiency of the classical FEL.
Both quantum parameters, the more fundamental $\alpha$ and the coupling strength quantifying $\bar\rho$, have their respective fields of application to which they are more natural.
Since this work focuses on the experimental realization of a Quantum FEL we adopt in the following mainly the notation of Ref.~\cite{boni06} in terms of the parameter $\bar{\rho}$.

\subsection{Interaction Length}

The quantum regime of the FEL is defined by the two-level behavior of the momentum states $p=q/2$ and $p=-q/2$. From quantum optics it is known that the time evolution of such a system is characterized by Rabi oscillations~\cite{schleich}. Hence, the probabilities $P_{p=q/2}$ and $P_{p=-q/2}$ for the electron to be in the excited state $q/2$ and the in ground state $-q/2$ of the two-level system are given by the simple relations~\cite{Kling2015}
\ba
P_{p=q/2}&=\cos^2{\Omega t}\\
P_{p=-q/2}&=\sin^2{\Omega t}\,,
\ea
where $\Omega=g\sqrt{n_\text{r}}$ denotes the Rabi frequency.

Hence, $2\Omega$ defines the typical frequency scale of the time evolution in the quantum regime. The typical length scale is then simply found by the relation
\ba
L_\text{g}=\frac{c}{2\Omega}\,,
\ea
where we have identified $L_\text{g}$ as the gain length of a Quantum FEL.
In terms of $\bar{\rho}$ this gain length reads~\cite{boni06}
\ba\label{eq:qtheory_Lg}
L_\text{g}=\frac{\gamma^2}{k_\text{FEL}}\frac{1}{\underbrace{\rho \sqrt{\bar{\rho}}}_{\equiv \Gamma}}\, ,
\ea
where we have used $\alpha=\Omega/\omega_\text{r}$, as well as eq.~\eqref{eq:qtheory_one_to_one} and the definition eq.~\eqref{eq:qtheory_rhobardef} of $\bar{\rho}$. 
Compared to classical FEL theory we had to replace the FEL parameter $\rho$, which also quantifies the gain bandwidth of the classical FEL, by the  quantity $\Gamma=\rho\bar{\rho}^{1/2}$. 
Later on we can identify $\Gamma$ as the gain bandwidth of a Quantum FEL.

Motivated by the heuristic argument that $\Gamma$ equals $\rho$ in the classical limit $\bar{\rho}\to \infty$, $\Gamma$ is occasionally defined \cite{Bonifacio2005b,BONIFACIO2007} by the interpolation formula
\begin{equation}
\Gamma=\rho\sqrt{\frac{\bar{\rho}}{1+\bar{\rho}}}\, .
\label{eq::gainBandwidthInterpolate}
\end{equation}
Although \eqref{eq::gainBandwidthInterpolate} is correct in both the quantum and classical limits, it constitutes only a first estimate for the transition range in between.

\section{Experimental challenges}
\subsection{Objectives and significance of Quantum FELs}
\subsubsection{Basic requirements and goals}

For realizing a Quantum FEL in the hard x-ray range, as for any classical FEL, we lack highly reflective x-ray optics and thus rely on an external seed or the SASE process, which starts from the shot noise of an electron bunch combined with the spontaneous Compton emission characteristics as seed.
Thus and in contrast to the low-gain amplification of an (Q)FEL oscillator within a (laser) cavity, the interaction to attain full amplification until saturation takes place within a single interaction between an electron bunch and a laser undulator.

Both the initial electron bunch and laser pulse of a QFEL shall follow well-defined, but still classical distributions.
Specifically, laser pulses are not described by a single frequency but rather by a finite bandwidth.
Instead, the classical initial laser and electron distributions reflect that frequency $\omega$ and momentum $\vec{p}$ are not discrete, but continuous observables.
Measurable quantum effects emerge within the QFEL interaction only.

The primary aim of this analysis is not the realization of the smallest possible natural bandwidth $L_\text{bunch}/\lambda_\text{FEL}$ of a QFEL, but instead the easier goal of significantly surpassing the bandwidth of a classical FEL $\Delta\lambda_\text{FEL}/\lambda_\text{FEL}=2\rho$. The measurable characteristics of a QFEL are: First, the decreased bandwidth $\Delta\lambda_\text{FEL}/\lambda_\text{FEL}=2\rho\sqrt{\bar{\rho}}$, which secondly, ideally consists only of a single spectral spike -- sometimes referred to as ``quantum purification'' \cite{Bonifacio2005b,bonifacio-basis,Bonifacio2017}. 
Finally, each electron emits at maximum one photon such that the resulting photon number $N_\text{QFEL,max}=N_\text{el}$ could in principle significantly surpass the expected photon number of the corresponding classical regime $N_\text{FEL,max}=\bar{\rho}N_\text{el}$.

Another goal is coherence according the definition \cite{Hecht2002} in classical optics which is also applied for classical FELs. 
Longitudinal coherence in QFELs diminishes when the electron bunch duration exceeds the quantum cooperation length $L_\text{c}$.
In other words, causally separated regions within a QFEL emit radiation within the same target bandwidth, but feature different relative phases, which leads to spectral interference patterns within the remaining ``SASE-spike''.
While good longitudinal coherence is a highly desirable trait, sizable degree of transverse coherence is critical to virtually all FEL applications, because the degree of coherence is related to the ability to focus and image the beam, as well as to perform subsequent interferometric or holographic techniques.
Thus, for a QFEL to be experimentally useful, we require a minimum x-ray photon number $>\num{e+5}$ with superior small bandwidth, while being longitudinally and transversally coherent.

\subsubsection{Significance of Quantum FELs for experiments and future applications}

The primary motivation for QFEL physics with regard to future experiments and potential applications lies in producing highly coherent pulses at hard x-rays from an ideally compact light source.
Long coherence lengths up to several \si{\um} significantly surpass the coherence lengths of existing hard x-ray lasers.
Beyond these applications there is also a fundamental interest in QFELs as a quantum system, in which a Quantum SASE FEL with its photon statistics is the natural extension of a SASE FEL with significant quantum mechanical recoil.
Even for classical FELs with $\bar{\rho}\gg 1$ quantum effects may become relevant, since in these regimes the position uncertainty of single electrons can already become larger than the radiated wavelength \cite{anisimov2018} during interaction.
Therefore, a quantum physics description is necessary to understand the FEL evolution, even if only to confirm that there is indeed no observable effect.

The biggest drawback of a QFEL besides its strict experimental requirements appears to be the relatively small x-ray photon number, which at best equals the number of electrons. While a QFEL would have a photon yield of \num{e+5} to \num{e+10} per pulse, an XFEL-type source already features \num{e+12} photons per pulse \cite{Yabashi2017, Pellegrini2016}.
Hence one could put forward the argument, that the same spectral width can be attained by applying a narrow spectral bandpass filter.
However, this does not change the coherence time of the resulting x-ray pulse.
Instead of a well-defined Gaussian envelope in both time and spectral domain, one would end up with a shot-by-shot, randomly distributed train of pulses.
A Quantum FEL, on the other hand, can produce small x-ray-laser bandwidths in the spectral domain and simultaneously well-defined, ultrashort temporal pulse characteristics.
In the best case, a QFEL based on an ultrashort electron bunch would also yield an x-ray pulse of the same length and not a laser envelope consisting of numerous, causally independent intensity spikes.

Thus, the central value of hard x-ray radiation from a QFEL for experiments and applications would be a significantly improved longitudinal coherence length, potentially exceeding the coherence length of classical FELs by one order of magnitude.
Combined with its well-defined, ultra-short temporal x-ray laser structure this feature would enable novel applications such as phase-referenced (i.e. interferometric and holographic), time-resolved studies of the structure of matter. The field of warm-dense matter physics is one example, where phase-sensitive diagnostics could identify exotic states of matter undergoing ultrafast instability dynamics and rapid phase transitions in highly overdense plasmas.

Analogously to the development of ultrashort, high-power, optical lasers, where an initial low-power, but high-quality seed is key to an entire amplification chain, an ultrashort QFEL x-ray laser pulse of good pulse quality but limited photon number could serve as a seed for subsequent amplification processes.

From an experimental perspective, QFELs would most-likely be driven by compact optical laser undulators which require much lower electron energies.
Compared to kilometer long linear accelerators and undulators, such designs would be much more compact. In contrast to classical SASE, one exploits that once the quantum regime is reached, the number of emitted photons depends only on the number of electrons available and not on the electron energy.

\subsection{Non-ideal effects}
\subsubsection{Introduction to non-ideal effects}

A high quantum mechanical recoil, quantified by $\alpha=\bar{\rho}^{3/2} \ll 1$, is not the only fundamental constraint to realize the quantum regime of the FEL.
If we want to avoid that the discreteness of the momentum steps is washed out we require the initial momentum spread $\Delta p$ of the electron beam in the average rest frame to be smaller than the separation $q$ of the momentum levels yielding $\Delta p < q$~\cite{Kling2015}.
In the laboratory frame we then obtain
\begin{align}
\label{eq:qtheory_ladder}
\frac{\Delta \gamma}{\gamma} &<\frac{\hbar k_\text{FEL}}{\gamma m c}=\frac{\rho}{\bar{\rho}}\,.
\end{align}
Due to $\lambda_\text{FEL}\propto\gamma^2$ this requirement for the energy spread translates to a
maximum linewdith
\begin{align}
\frac{\Delta \lambda_\text{FEL}}{\lambda_\text{FEL}} &<2\frac{\hbar k_\text{FEL}}{\gamma m c}=2\frac{\rho}{\bar{\rho}}\,
\label{eq:ladder_two}
\end{align}
for the FEL radiation.

We emphasize that Eqs.~\eqref{eq:qtheory_ladder},~\eqref{eq:ladder_two} only represent limits at which quantum effects become visible in the FEL.
However, they do not make any statement about the parameter space in which these can be efficiently exploited in order to operate a Quantum FEL.
in the following we show that the parameter limits allowing for (efficient) operation of a Quantum FEL are closely related to its gain bandwidth.

The dynamics of the electron in the quantum regime is characterized by sharp resonances due to energy-momentum conservation; that is the electron jumps from excited to ground state, only if in the average rest frame its initial momentum exactly equals the resonant one $p=q/2$.
Since the interaction time is finite these resonances are broadened and interaction also takes place for momenta which are slightly off-resonant.

We can quantify this broadening of the resonance by considering energy-time uncertainty
\ba\label{eq:qtheory_Et}
\updelta E ~\updelta t \geq \frac{\hbar}{2}\,,
\ea
where $\updelta E$ and $\updelta t$ denote the uncertainties in energy and time, respectively.
We estimate  $\updelta t$ by the  typical interaction time characterized by the Rabi frequency
$\Omega$ yielding  $\updelta t=1/(2\Omega)$.
For the energy of the electron we simply consider the free energy
$E=p^2/2m$ which yields for a resonant electron with $p=q/2$ the expression
\ba\label{eq:qtheory_dE}
\updelta E\cong\left(\frac
{\partial E}{\partial p}\right)_{p=\frac{q}{2}}\!\cdot\updelta p=\frac{1}{2}\frac{q}{m}\updelta p
\ea
in terms of a momentum uncertainty $\updelta p$ .

With the help of Eq.~\eqref{eq:qtheory_dE} and by taking the lower bound of Eq.~\eqref{eq:qtheory_Et} into account we finally obtain  the relation
\ba\label{eq:qtheory_broadening}
\frac{\updelta p}{q}= \alpha
\ea
for the broadening $\updelta p$ of the resonance in momentum space, where we have employed $\alpha=\Omega/\omega_\text{rec}$.
A more rigorous derivation yields the similar result $\updelta p=2\alpha q$.
If the initial momentum spread $\Delta p$ of the electron beam is smaller than this broadening each electron participates in the interaction and contributes to the gain of the FEL.
For an electron beam which possesses a momentum spread that exceeds the constraint given in Eq.~\eqref{eq:qtheory_broadening}, however, only a fraction of the electrons interacts with the fields, which is known as velocity selectivity in cold atom physics~\cite{giltner,giese,szigeti}.
In this case, the gain is reduced and thus we identify an offset $\updelta p=\alpha q$ as the gain bandwidth of a Quantum FEL.
Since $\alpha \ll 1$ in the quantum regime, we, moreover, recognize that this requirement due to velocity selectivity is stronger than the fundamental constraint $\Delta p< q$.

In terms of the laboratory frame we obtain the condition
\ba
\label{eq::QFELEnergyWidth}
\frac{\Delta \gamma}{\gamma}<\sqrt{\bar{\rho}}\rho = \Gamma\,,
\ea
for the relative energy spread of the electron beam, where we have used Eqs.~\eqref{eq:qtheory_one_to_one} and~\eqref{eq:qtheory_ladder} as well as the definitions, Eqs.~\eqref{eq:qtheory_rhobardef} and~\eqref{eq:qtheory_Lg}, of $\bar{\rho}$ and $\Gamma$, respectively.
This heuristically motivated result is consistent with existing literature on Quantum FELs~\cite{Bonifacio2017}.
Since $\Gamma$ denotes the maximally allowed energy spread for which amplification occurs, we identify this quantity as the gain bandwidth of a Quantum FEL.

By now there are two requirements on the electron energy spread.
One for the emergence of quantum effects in the interaction and the other for efficiently driving a QFEL, eqs.\ \eqref{eq:qtheory_ladder} and \eqref{eq::QFELEnergyWidth} respectively.
The first of these is a necessary condition and is automatically fulfilled for quantum regimes $\bar{\rho}\leq 0.62$ if the latter condition is fulfilled.
For regimes with a larger quantum parameter the latter condition is not sufficient to analyze the requirements for QFEL operation.
Electrons violating condition \eqref{eq:qtheory_ladder} do not only reduce the ability to drive the QFEL instability, but can actively disrupt the process by undergoing competing transitions.

Three-dimensional non-ideal effects are best systematically analyzed by starting from the relative bandwidth of the amplified radiation in a Quantum FEL, which in contrast to a classical FEL is
\begin{align}
\frac{\Delta\lambda_\text{FEL}}/{\lambda_\text{FEL}} &\leq 2\Gamma = 2\rho\sqrt{\frac{\bar{\rho}}{1+\bar{\rho}}}\\
                                &\simeq 2\rho\sqrt{\bar{\rho}}\,\text{, for $\bar{\rho}\ll 1$.}
\label{eq::QFELRadBandwidth}
\end{align}
This bandwidth can then be compared to the kinematics of the fundamental scattering process, which here can be well described by the classical Thomson formula
\begin{equation}
\lambda_\text{FEL} = \frac{\lambda_u\cdot\left( 1 + a_0^2/2 + \gamma^2\theta^2 \right)}{2\gamma^2}\,,
\label{eq::ThomsonFormular}
\end{equation}
with
\begin{equation}
\lambda_u=\frac{\lambda_0}{1 - \beta_0\cos{\phi}}\, .
\end{equation}
describing an effective one-dimensional undulator wavelength for arbitrary interaction angles $\phi$.
The interaction angle is enclosed by the directions of propagation of the electron bunch and the laser pulse providing the optical undulator field.
Above eqs. feature the wavelength dependence on the observation angle $\theta$ with respect to the electron direction of propagation, laser wavelength $\lambda_0$ and the electron energy and velocity, which is expressed in terms of the speed of light, $\gamma$ and $\beta_0$ respectively \cite{Ride1995}.

The dimensionless laser strength
\begin{align}
a_0 & = \frac{e E_0}{m c \omega_0}\\
    & \simeq \num{0.8493e-9}\lambda_0[\si{\um}]I_0^{1/2}[\si{W/cm^2}]\, ,
\label{eq::a0HandyDef}
\end{align}
where $E_0$ denotes the laser electric field, $I_0$ its intensity and $\omega_0$ its central angular frequency, characterizes the transition from sub-relativistic $a_0\ll 1$ to relativistic $a_0\ge 1$ quiver velocities of electrons within the laser field. For optical undulators $a_0$ identically replaces the usual dimensionless undulator parameter $K=e B_0\lambda_u/2\pi m c$ of magnetic undulators. Physically, this analogy exists, because undulators were designed to mimic at a very good approximation an incident electromagnetic wave with respect to a relativistic electron beam. Therefore, on the level of fundamental processes one can easily interchange undulator and Thomson scattering theory. In non-ideal laser and electrons beams, the quantities $a_0$, $\lambda_0$, $\phi$ and $\gamma$ are neither uniform in space nor constant over time. Essentially, in order to contribute to the same QFEL transition, the radiation of every electron needs to remain both inside the interaction region and within the gain bandwidth eq.~\eqref{eq::QFELRadBandwidth}.

However, this simple picture quickly becomes more complicated, because electron momentum and laser frequency are both continuous and not discrete observables.
Thus one has to consider not one single line transition of some specified relative spectral width $2\Gamma$, but a whole range of possible transitions between two bands of momentum states driven by a laser of finite and continuous frequency bandwidth.
All these transitions need to remain within the aforementioned $2\Gamma$ bandwidth.
Accordingly, the best case result for Dirac-delta-type distributions of electron and laser spectra are $N_\text{phot,max}\simeq N_\text{el}$ QFEL laser photons with at minimum the natural bandwidth
\begin{equation}
\frac{\Delta\lambda_\text{FEL,min}}{\lambda_\text{FEL}}=\frac{L_\text{bunch}}{c}\, .
\label{eq::QFELMinimumBandwidth}
\end{equation}
With finite spectral widths as in any real-world experiment, this minimum bandwidth increases according to the electron momentum distribution and laser bandwidth until the QFEL bandwidth criterion eq.~\eqref{eq::QFELRadBandwidth} cannot be met anymore and prevents the QFEL interaction.

In the following we go through the different properties of electron bunch, laser pulse, interaction geometry, space-charge and spontaneous emission to identify critical challenges for experimental realizations of QFELs. Here we adopt standard FEL nomenclature for the 1D-FEL power gain length of a monoenergetic beam \cite{boni06}\footnote{Note, that the QFEL dispersion relation is not a cubic equation as for the classical 1D-FEL \cite{Pellegrini2016,Huang2007,Barletta1993}. Thus the gain length of the classical FEL for $\bar{\rho\gg 1}$ is by a factor of $\sqrt{3}$ smaller than for a QFEL. Consistently, the required number of gain lengths $N_\text{g}$ until saturation for a classical FEL is also by a factor of $\sqrt{3}$ higher than for a QFEL, hence resulting in a similar saturation length for the classical and the quantum regime.}
\begin{equation}
L_\text{g}=\frac{\lambda_u}{4\pi\Gamma}\, ,
\label{eq::QFELGainLength}
\end{equation}
as well as the cooperation length
\begin{equation}
L_\text{c}=\frac{\lambda_\text{FEL}}{4\pi\Gamma}\,
\label{eq::QFELCooperationLength}
\end{equation}
in order to be more consistent with respect to standard literature on classical FELs. For Quantum FELs $\rho$ is replaced here with $\Gamma$. Compared to some previous work on QFELs \cite{Bonifacio2005b,BONIFACIO2007} these definitions are smaller by a factor of $\sqrt{3}$. Thus, the expected saturation length
\begin{equation}
L_\text{sat}\simeq 10 L_g \sim \frac{\lambda_u}{\Gamma}\, .
\label{eq::QFELSaturationLength}
\end{equation}
accordingly increases.

\subsubsection{Challenges due to required electron bunch properties}

Electron energy spread and transverse emittance of the electron beam both enlarge the radiation bandwidth. They are strict constraints for experimentally reaching the quantum regime of the FEL.
Especially if the condition for isolating a single momentum transition \eqref{eq:ladder_two} is violated, discrete FEL frequency-bands ensured by the nonlinear energy-momentum relation $\gamma(p)=\sqrt{1+(p/mc)^2}$ cease to exist. Equation~\eqref{eq::ThomsonFormular} relates the energy spread to a spread in radiated wavelength $2\Delta\gamma/\gamma=\Delta\lambda/\lambda$, thus the condition on the electron bunch energy spread is given by
\begin{equation}
\frac{\Delta\gamma}{\gamma}\leq\Gamma=\rho\sqrt{\frac{\bar{\rho}}{1+\bar{\rho}}}
= \frac{\lambda_c}{\gamma\lambda_\text{FEL}}\sqrt{\frac{\bar{\rho}^3}{1+\bar{\rho}}}\, ,
\label{eq::energySpreadConstraint}
\end{equation}
which we obtained earlier, cf.\ \eqref{eq::QFELEnergyWidth}, but now we applied the interpolation formula \eqref{eq::gainBandwidthInterpolate} for $\Gamma$.
The strong energy dependence $\Delta\gamma/\gamma\propto \gamma^{-1}$ shows that conventional undulators are at a significant disadvantage for QFELs. According to eq.~\eqref{eq::ThomsonFormular}, where $\lambda_\text{FEL}\simeq \lambda_u/2\gamma^2$, a cm-long undulator wavelength $\lambda_u=\SI{1}{cm}$ of a magnetic undulator requires such high electron energies (\SI{3.6}{GeV}) that the relative energy spread according to eq.~\eqref{eq::energySpreadConstraint} for $\bar{\rho}=0.2$ needs to fall below a limit of \num{2.8e-7}. While the absolute energy spread is the same as for optical undulator with smaller $\lambda_u$, reaching such low relative spreads after acceleration to higher energies, in which each acceleration cavity increases energy spread, is currently beyond the state-of-the-art. Hence, more compact optical undulators with $\lambda_u$ in the range of \si{\um} to \SI{100}{\um} do not only lead to more compact setups, but significantly reduce technical demands on the accelerator to provide extraordinary small relative electron energy spreads.

The normalized transverse emittance $\varepsilon_n$ of an electron bunch combined with the (rms) electron beam radius $\sigma_\text{e}$ is a measure of the electron beam divergence. According to eq.~\eqref{eq::ThomsonFormular} and \cite{Steiniger2014_3} the condition for the maximum allowed divergence angle reads
\begin{equation}
\gamma^2\delta\phi^2\leq 2\Gamma
\label{eq::divergenceConstraint}
\end{equation}
and thus the normalized emittance limit is
\begin{equation}
\varepsilon_n\leq\sigma_\text{e}\sqrt{2\Gamma}\, .
\label{eq::emittanceConstraint}
\end{equation}

The electron density of a divergent electron beam also reduces during propagation which diminishes the electron--radiation coupling.
The laser and electron pulse overlap over the interaction distance $L_\text{int}$ is reduced, too.
Both effects lead to an additional criterion on the normalized electron bunch emittance
\begin{equation}
\varepsilon_n \leq \sigma_\text{e}^2\gamma/L_\text{int}\, ,
\label{eq::defocusingContraint}
\end{equation}
where we require the interaction length to remain shorter than twice the $\beta^\star = \gamma\sigma_\text{e}^2/\varepsilon_n$ of the electron bunch, which denotes the $\beta$-function in the middle of the interaction at the focus $\beta(s)=\beta^\star+s^2/\beta^\star$ and represents the characteristic defocusing distance. This defocusing condition is of high practical relevance, since optical FELs (OFELs) do not \emph{a priori} provide focus guiding of the electron beam as in magnetic undulators, where the increasing magnetic field off-axis can confine an electron beam. Although optical undulators could at least in principle achieve the same through a ponderomotive potential of carefully engineered intensity gradients in the laser pulse, the technical implementation is extremely challenging.
For typical QFEL electron beam parameters beyond \SI{30}{MeV} energy, $\varepsilon_n=\SI{0.01}{mm.mrad}$ normalized emittance and $\sigma_e=\SI{2.0}{\um}$, one typically arrives  interaction lengths at the centimeter scale(here: $2\beta^\star=\SI{4.8}{cm}$).

Finally, the electron bunch charge $Q$, rms cross-sectonal radius $\sigma_\text{e}$ and pulse length $\tau_\text{e}$, determine the bunch mean electron density $n_\text{e}$ and hence the classical FEL coupling strength $\rho$, as well as the Quantum FEL coupling strength
\begin{align}
\Gamma\propto\rho &=  \left[ \frac{a_0^2 f_\mathrm{B}^2 \Omega_\mathrm{p}^2}
        {32 \gamma_0^3 c^2 k_0^2 (1-\beta_0\cos\phi)^2}
    \right]^{1/3}\label{eq::defineRho}\\
    &= \left[ \frac{1}{16 \gamma_0^3} \frac{I}{I_\mathrm{A}} \left( \frac{\lambda_0 a_0 f_\mathrm{B}}{2 \pi \sigma_\text{e} (1-\beta_0\cos\phi)} \right)^2\right]^{1/3}\\
    &\propto Q^{1/3}\tau_\text{e}^{-1/3}\sigma_\text{e}^{-2/3}\, ,
\end{align}
with the non-relativistic plasma frequency $\Omega_\mathrm{p}^2 = e^2 n_\text{e}/\varepsilon_0 m$, $f_\mathrm{B} = [ J_0( \chi ) - J_1(\chi)]$ with $\chi = a_0^2 / (2+a_0^2)$, the Alfv\'{e}n current $I_\mathrm{A}=4\pi\varepsilon_0 mc^3/e\approx \SI{17}{kA}$, and the electron beam peak current $I$.

In order to enter the quantum regime of an FEL extraordinary high-quality electron guns are needed to realize the required small normalized emittances $\varepsilon_n$ at a scale below $\varepsilon < \SI{0.1}{mm.mrad}$.
Developing such guns is an active field of research within accelerator physics.
However, when assuming only minor extrapolations beyond the current state-of-the art, one of the promising methods appears to be going towards small bunch charges in order to isolate electron bunches with low-emittances \cite{Rosenzweig1995,Zhou1999, Ding2009, Bartnik2015} at relatively high electron densities.
While this sacrifices final photon flux, such an approach could potentially meet the strict electron requirements towards QFEL proof of principle experiments \cite{Bonifacio2017}.

\subsubsection{Challenges due to required laser undulator properties}

In the deep quantum regime $\bar{\rho}\ll 1$, the undulator wavelength is chosen such that $\rho$ and thus $\Gamma\simeq\rho\sqrt{\bar{\rho}}$ are being maximized to reduce the engineering and economic challenges towards an experimental realization.

The duration of the entire laser-electron interaction is $L_\text{sat}/c$. In head-on, colliding geometries, where laser and electron beam foci overlap and propagation directions are oriented \SI{180}{\degree} with respect to each other, the laser pulse duration $\tau_0$ is required to be longer than
\begin{equation}
\tau_0 > 2\cdot L_\text{sat}/c\,.
\label{eq::laserDurationConstraint}
\end{equation}
For this reason most QFEL scenarios with \si{mm} to \si{cm} long interaction lengths require a laser pulse duration in the \si{ps} to \si{ns}-range. Later in this article we will show how this constraint can be relaxed.

Due to the onset of nonlinear Thomson scattering at $a_0$ approaching unity, intensity variations also lead to variations of the scattered wavelength by the $(1 + a_0^2/2)$ redshifting-factor in eq.\eqref{eq::ThomsonFormular}. Hence the upper limit on intensity variations due to the wavelength shift for $a_0 < 1$ is
\begin{equation}
\frac{\delta I_0}{I_0}\leq 4\Gamma\frac{1+a_0^2/2}{a_0^2}\, .
\label{eq::intensityVariationConstraint}
\end{equation}
In practice this is a strong limit for high laser intensities, since it is technically extremely challenging to provide spatially and temporally uniform laser beams of laser strengths $a_0$ of unity and beyond, where required intensity variations would often be in the sub-percent range.

The variation in laser intensity during the propagation of the electron bunch caused by the laser spatial envelope gives rise to a ponderomotive force on the electrons. Its strength is $\bm{F}=-mc^2 \nabla a_0^2/4\gamma$ \cite{Gibbon2005}, provided that $a_0\ll\gamma$. Through intensity gradients within the laser pulse, the ponderomotive force causes a drift in the electron motion away from the center of the bunch. To minimize the resulting error angle $\delta\phi$, the relative intensity variation $\delta I_0/I_0 = 2\delta a_0/a_0$ must be kept small. In order to derive an approximate limit on the allowed intensity variation within the laser pulse, we assume a constant ponderomotive force over the entire interaction time and approximate the gradient linearly with
$\nabla a_0^2 \approx \sqrt{2/\pi} \delta a_0^2 / \sigma_\text{e}$. Hence the deflection angle $\theta_\text{pond} = \delta a_0^2 L_\text{int} / \sqrt{8 \pi} \gamma^2 \sigma_\text{e} $ combined with condition eq.~\eqref{eq::divergenceConstraint} yields the intensity variation limit of the ponderomotive force
\begin{equation}
    \frac{\delta I_0}{I_0} \le \frac{4}{a_0}
        \left( \frac{\pi \rho \sigma_\text{e}^2 \gamma^2}{L_\text{int}^2} \right)^{1/4}\, .
\label{eq::pondForceConstraint}
\end{equation}

Typically intensity variations within each pulse of several percents can be obtained in modern high-power laser systems. Beyond improving the optics and material quality, as well as thermal control and vibration stability, there exists little work on active filtering and control of small scale intensity variations.
Existing homogenizing filters as used in industry applications \cite{Voelkel2008} are unsuitable for driving QFELs, since homogenizers that reduce local intensity noise within a laser pulse strongly reduce its spatial coherence, too.

\subsubsection{Challenges due to coherence properties}

A high degree of coherence $\langle \bm{E}(\bm{r_A},t) \bm{E}^\star(\bm{r_B},t+\tau) \rangle > 0$ in resulting FEL x-ray beams is essential to almost all applications. Spatially coherent FEL beams show planar phase fronts, which makes it possible to image and focus the beam. Longitudinally (or temporally) coherent FEL beams feature an extended coherence length over which a fixed phase relation exists that can be used in experiments as prior knowledge. For interferometric or diffraction-based techniques, both spatial and temporal coherence usually has to exceed the dimensions of the target under investigation.

For identifying the maximal achievable domain of full coherence both in classical and Quantum FELs, it is useful to identify the volumes within the electron beam that are causally connected by the entire FEL interaction. Causally disconnected regions in an unseeded FEL, can generate FEL radiation at the same wavelength, but do so with random phase shifts with respect to one another.

Longitudinally, this dimension is defined by the so called cooperation length, i.e. the total slippage length between the x-ray pulse and electrons.
According to the FEL resonance condition, the emitted light slips over the electrons by one FEL wavelength for every electron oscillation period which defines the total slippage length as the radiation wavelength times the number of undulator periods until saturation.
For a QFEL we assume that this cooperation length is approximately given by
\begin{equation}
  L_c = \frac{\lambda_\text{FEL}}{4 \pi\Gamma}\,\text{, with } \Gamma = \rho \sqrt{\frac{\bar{\rho}}{1+\bar{\rho}}}
\end{equation}
in analogy to classical FEL theory.
Compared to eq.\ \eqref{eq::QFELCooperationLength} the interpolated expression for $\Gamma$ is used now.

Transversally, the degree of coherence is not as straightforward to identify as for longitudinal coherence, primarily because the interplay between electron beam and FEL laser usually requires more extensive numeric modelling.
For build-up of transverse coherence there needs to be some interaction between distinct regions of the beams transverse cross-sections in order to permit synchronization, which is homogenization, of the FEL beams transverse phase front.
This does not occur, if both FEL and electron beam are perfectly collinear.
However, diffraction of finite-size FEL beams, as well as divergence of electron beams with non-zero emittance and space-charge all lead to variations of the transverse beam cross-sections and FEL radiation modes.
Assuming equal diameters, one usually compares divergence of electron and FEL beams by their respective emittances $\varepsilon_n/\gamma$ and $\varepsilon_r=4\pi/\lambda_\text{FEL}$.

Furthermore we assume that the electron and the FEL photon beam fulfill the emittance eq.~\eqref{eq::defocusingContraint} and diffraction criteria, $L_\text{g}<z_\text{r,FEL}=4\pi\sigma_e^2/\lambda_\text{FEL}$ \cite{Steiniger2014_3} respectively, to avoid excessive losses in the FEL gain.
Within these constraints, a diverging FEL beam imprints the local FEL phase onto the local phase of the modulated electron beam and vice versa. 
While the existence of beam divergence effectively synchronizes the phase front over the course of a number of gain lengths, this competition of transverse modes during amplification tends to be more effective if FEL and electron beam divergences are matched.
Thereby the respective interaction duration between local electrons and FEL photons is maximized.

Qualitatively, there are two extremes for radiation emittance: $\varepsilon_r \gg \varepsilon_n$ leads to transversal coherence, but strongly increases the FEL gain length.
Here a divergent FEL beam reduces the FEL field within the electron bunch and thus the FEL gain.
In the opposite limit $\varepsilon_r \ll \varepsilon_n$ maintains the shortest FEL gain length, but diminishes or eliminates transverse coherence since a low-divergent FEL beam does not reach all areas of the electron beam crosssection.
This is equivalent to independent Quantum SASE FELs in the same beam without mutual phase-lock and hence without transverse coherence.

Quantitatively, the degree of transverse coherence of FEL radiation is often characterized by the transverse coherence parameter \cite{Saldin2008,Saldin2008_2}
\begin{equation}
\hat{\varepsilon} = 2\pi\varepsilon_n/\gamma\lambda_\text{FEL}\, .
\end{equation}
The parameter $\hat{\varepsilon}$ matches the divergence of the FEL radiation and the electron beam such that the degree of transverse coherence is best at $\hat{\varepsilon} \approxeq 1$ and useful transverse coherence is typically available in a rather wide range of $\hat{\varepsilon} = 0.5 - 10$.

For a head-on laser geometry, where laser and electron collide collinearly at $\SI{180}{\degree}$ interaction angle, at \SI{800}{nm} wavelength and a permissive transverse coherence parameter target of $\hat{\varepsilon}=10$, the required normalized beam emittance for a QFEL at $\lambda_\text{FEL}=\SI{1}{\angstrom}$ is $\varepsilon_n=\SI{7.1e-3}{mm.mrad}$, which becomes $\varepsilon_n=\SI{7.1e-4}{mm.mrad}\propto\hat{\varepsilon}$ for perfect emittance matching $\hat{\varepsilon}=1$. This shows, that full transverse coherence can become prohibitively difficult.
This problem can be avoided by using drive lasers at longer wavelengths, such as $\text{CO}_2$ lasers at \SI{10.6}{\um}, or optical lasers in a Traveling-Wave Thomson-Scattering (TWTS) geometry \cite{Steiniger2016}.

Beyond the general guideline for $\hat{\varepsilon}$ a numerical modeling is essential for predicting transverse coherence and further radiation field properties in classical FELs.
Today there is no conclusive framework available for numerical simulations of the interaction between electrons, undulator field, radiation field as well as the evolution of the fields (e.g.\ laser defocusing and transverse mode competition respectively) in Quantum FELs starting from a realistic electron bunch distribution.
Yet first approaches using a quantum fluid approach exist~\cite{boni_wigner}.

Despite aforementioned limitations of current theoretical works, one expects that these general $\hat{\varepsilon}$ scalings hold also for Quantum SASE FELs.
In essence both the degree of longitudinal and transverse coherence of a SASE FEL primarily depend on the ability of the FEL field to causally connect spatially separated regions within the electron bunch and that the effect of slippage and electron beam divergence do not fundamentally change in the quantum regime.
Therefore we adopt the classical coherence requirements for Quantum SASE FELs
\begin{equation}
\varepsilon_n= \hat{\varepsilon} \frac{\gamma \lambda_\text{FEL}}{2\pi}\, ,
\label{eq::QFEL_tranverseCoherence}
\end{equation}
where the dimensionless transverse coherence parameter $\hat{\varepsilon}$ needs to be within the approximate range $\hat{\varepsilon}\in [0.5 \ldots 10]$ in order to obtain at least partial transverse coherence.

\subsection{Quantum FEL Interaction Geometry}
\subsubsection{Challenges due to the interaction region}

In the usual head-on scattering geometry, which is a collinearly, colliding configuration, the interaction length does not only need to be shorter than twice the $\beta^\star$ of the elecron beam, cf.\ \eqref{eq::defocusingContraint}, it is also required to remain shorter than twice the laser Rayleigh length
\begin{align}
  L_\text{int} &< 2 z_r\, ,\\
  \text{with } z_r &= \frac{\pi w_0^2}{\lambda_0}\, ,
\label{eq::QFEL_RayleighConstraint}
\end{align}
which denotes the distance from the laser waist where the width of the beam is $\sqrt{2}$ larger compared to $w_0$ at the waist and on-axis intensity one half of the peak intensity. For a laser focal width $w_0=\SI{10}{\um}$ at \SI{800}{nm} wavelength, chosen to be considerable larger than the electron bunch radius $\sigma_\text{e}=\SI{2}{\um}$, the available interaction distance $2 z_r=\SI{0.078}{cm}$ is much smaller than $2\beta^\star=\SI{4.8}{cm}$.

Moreover, the change in laser intensity due to defocusing, the Gouy-phase, the intensity variation originating from the transverse laser profile, as well as the laser curvature further limit the useful interaction volume.

According to the field amplitude scaling $a_0\propto(1 + (z/z_r)^2)^{-1/2}$ of a collimated Gaussian beam with propagation distance z from the waist and eq.~\eqref{eq::ThomsonFormular}, the resulting relative deviation in FEL wavelength during interaction and corresponding useful range in $(z/z_r)$ for a QFEL with $2\Gamma$ bandwidth are
\begin{align}
\frac{\Delta\lambda_\text{FEL,defocus}}{\lambda_\text{FEL}} &= 1-\frac{(1+(f_{a_0} a_0)^2/2)}{1+a_0^2/2} \text{ , with }f_{a_0}=(1+(z/z_r)^2)^{-1/2}\\
\frac{\Delta\lambda_\text{FEL,defocus}}{\lambda_\text{FEL}} &= \frac{a_0^2 (z/z_r)^2}{4+4(z/z_r)^2}\,\text{, leads to range}\\
\left(\frac{\Delta z}{z_r}\right)_\text{defocus} &= \pm \frac{2\sqrt{2\Gamma}}{\sqrt{a_0^2-8\Gamma}}\,\text{, satisfying $\Delta\lambda_\text{FEL}/\lambda_\text{FEL}\leq 2\Gamma$.}
\label{eq::QFELRayleighIntensityCriterion}
\end{align}
In case of cylindrical focusing, as for the Traveling-wave Thomson scattering geometry introduced later, the allowable distance doubles.

The Gouy phase leads to local shifts in laser wavelength experienced by the electrons passing through the focal region at the speed of light. The Gouy-phase $\Psi(z)=\arctan(z/z_r)$ modifies eq.~\eqref{eq::ThomsonFormular} according to
\begin{align}
\frac{\Delta\lambda_\text{FEL,Gouy}(z)}{\lambda_\text{FEL}} &= \frac{\Delta\omega_\text{0,Gouy}(z)}{\omega_0}=\left(\frac{\omega_\text{Gouy}(z=0)-\omega_\text{Gouy}(z)}{\omega_0}\right)\\
 &= \frac{1}{\omega_0}\left(\left.\frac{\partial \Psi(z=ct)}{\partial t}\right|_{t=0} - \left. \frac{\partial \Psi(z=ct)}{\partial t} \right|_{t\to z/c} \right)\\
 &= \frac{1}{k_0}\left(\left.\frac{\partial}{\partial z}\arctan(z/z_r)\right|_{z=0}-\frac{\partial}{\partial z}\arctan(z/z_r)\right)\, .
\end{align}
The resulting relative deviation in FEL wavelength and corresponding range in $(z/z_r)$ for a QFEL with maximum $2\Gamma$ bandwidth are
\begin{align}
\frac{\Delta\lambda_\text{FEL,Gouy}(z)}{\lambda_\text{FEL}} &= \frac{z^2}{k_0 z^2 z_r + k_0 z_r^3}\\
\left(\frac{\Delta z}{z_r}\right)_\text{Gouy} &= \pm \sqrt{\frac{k_0 z_r 2\Gamma}{1-k_0 z_r 2\Gamma}}\, .
\label{eq::QFELGouyPhaseCriterion}
\end{align}

Assuming a Gaussian shaped transverse laser mode with $a_0=a_\text{0,max}\exp{(-x^2/w_0^2)}$, eq.~\eqref{eq::intensityVariationConstraint} shows the wavelength shift and extent of the center region to be
\begin{align}
\frac{\Delta\lambda_\text{FEL,Gauss}}{\lambda_\text{FEL}} &= \frac{a_0^2}{2+a_0^2}(1-\exp(-2 x^2/w_0^2))\\
\left(\frac{\Delta x}{w_0}\right)_\text{Gauss} &= \pm \frac{1}{\sqrt{2}}\sqrt{\log{\frac{a_0^2}{a_0^2-4\Gamma-2a_0^2\Gamma}}}\, .
\label{eq::QFELGaussTransIntensity}
\end{align}

For the sake of completeness we note that wave curvature contributions from the focusing can alter the incident angle on individual electron by $\Delta\phi = \arctan(x/(z (1+(z_r/z)^2)))$. On axis ($x=0$) this effect typically becomes negligibly small, such that above interaction region constraints are more strict. This can be checked, when we insert $\Delta\phi$ in eq.~\eqref{eq::ThomsonFormular} and search for the maximum transverse extent of the interaction region at maximum wavefront curvature at $z=z_r$ that is permissible at the QFEL bandwidth $\Gamma$. For small angles $\Delta\phi\ll 1$ we arrive at the estimate
\begin{align}
\frac{\Delta\lambda_\text{FEL,curvature}}{\lambda_\text{FEL}} &= \frac{2}{1-\cos(\pi+\Delta\phi)}-1 \leq 2\Gamma\\
\left(\frac{\Delta x}{w_0}\right)_\text{curvature} &\simeq \pm 2\pi w_0 \sqrt{2\Gamma}(z/z_r+z_r/z)/\lambda_0\, .
\label{eq::QFELWaveFrontCurvature}
\end{align}

In an example scenario of an \SI{800}{nm} laser at $a_0=0.05$, focused to $w_0=\SI{10}{\um}$ with a Rayleigh length $z_r=\SI{393}{\um}$ for a QFEL with bandwidth $\Gamma=\num{5.0e-5}$, the Rayleigh-intensity criterion eq.~\eqref{eq::QFELRayleighIntensityCriterion} is $\Delta(z/z_r)=\pm 0.30$, the Gouy-phase criterion eq.~\eqref{eq::QFELGouyPhaseCriterion} is $\Delta(z/z_r)=\pm 0.67$, the transverse Gaussian intensity criterion is $\Delta(x/w_0)=\pm 0.20$ and the minimal radius according to the Rayleigh wavefront curvature criterion is $\Delta(x/w_0)=\pm 1.57$.

These results show that laser spot sizes need to be considerably larger than the corresponding electron beams.
In practice this leads to either prohibitively large requirements on the laser pulse energy or to extremely small radii for electron beams and thus to an even higher constraint on electron beam emittance, cf.\ eq.~\eqref{eq::emittanceConstraint}.
In essence, this is known as the Rayleigh limit.
Previous works \cite{BONIFACIO2007} impose this limit for finding suitable QFEL designs, therefore sacrificing much of the potential parameter space for QFEL configurations.
In the following section we briefly show that above Rayleigh-limit based constraints can be mostly avoided if a different interaction geometry is used.

\subsubsection{Traveling-wave laser geometries -- sidestepping the Rayleigh limit}

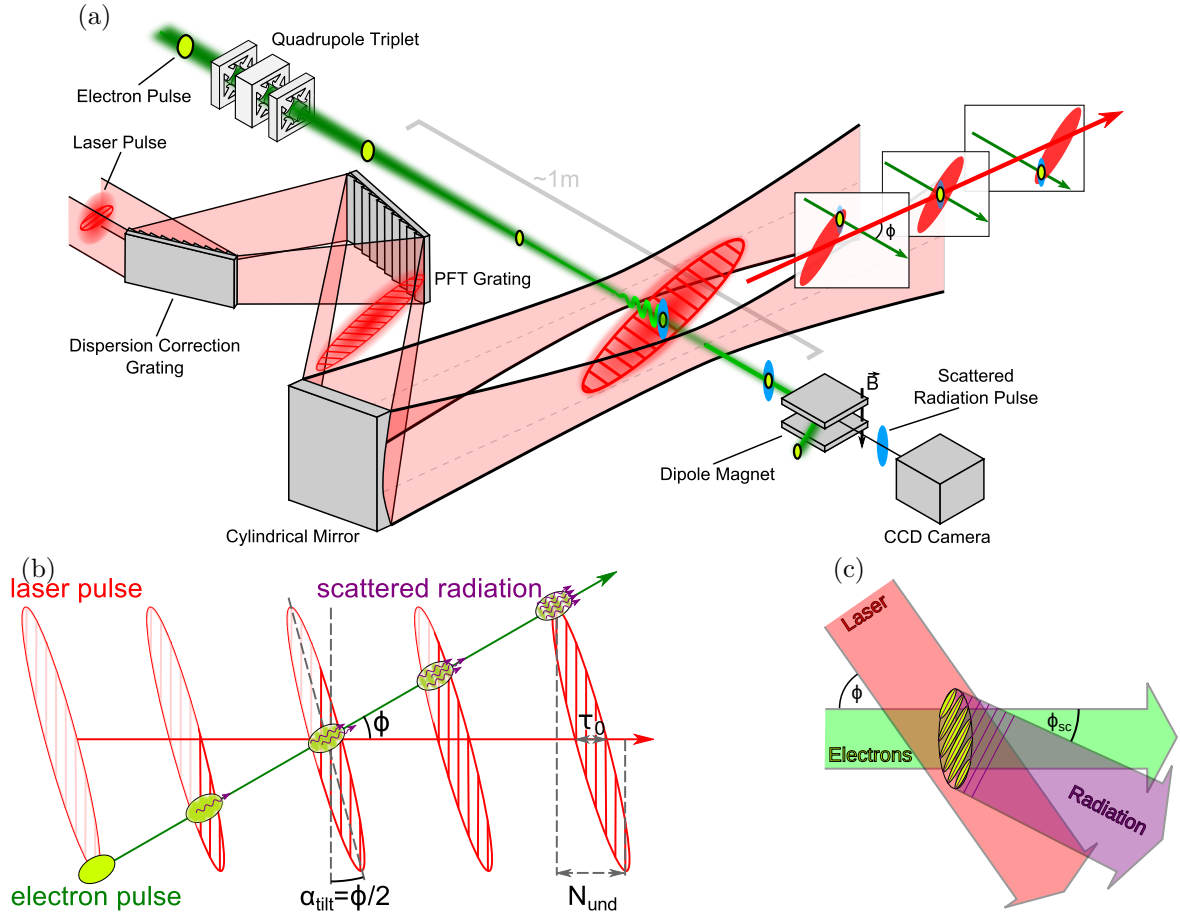
\begin{figure}[!t]
  \centering
  \begin{tikzpicture}
    \node[anchor=north west,inner sep=0] at (0,0) {\includegraphics[height=7.0cm]{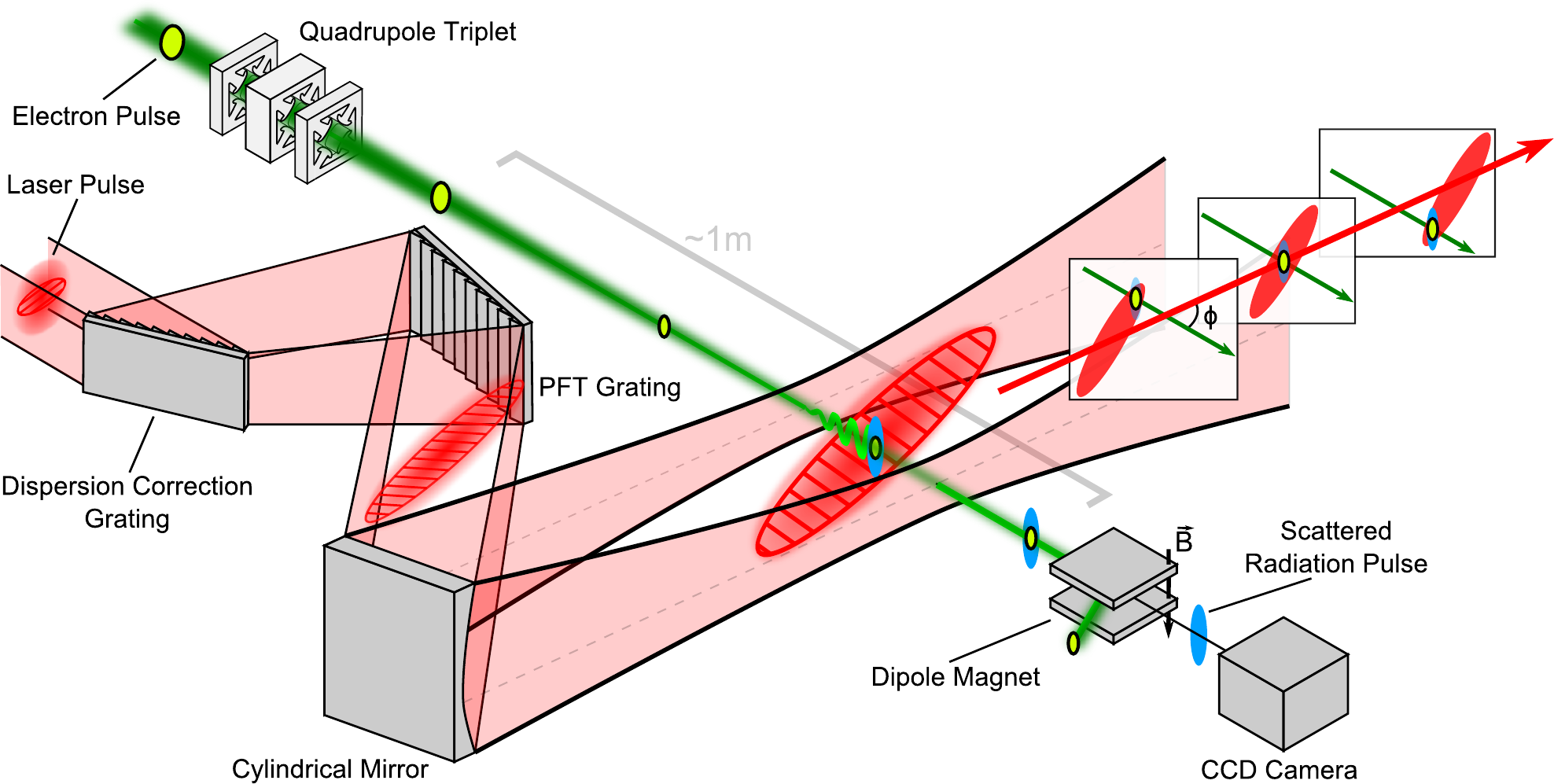}};
    \node[right,align=left] at (0.0,0.0) {(a)};
  \end{tikzpicture}\\
  \begin{tikzpicture}
    \node[anchor=north west,inner sep=0] at (0,0) {\includegraphics[height=4.5cm]{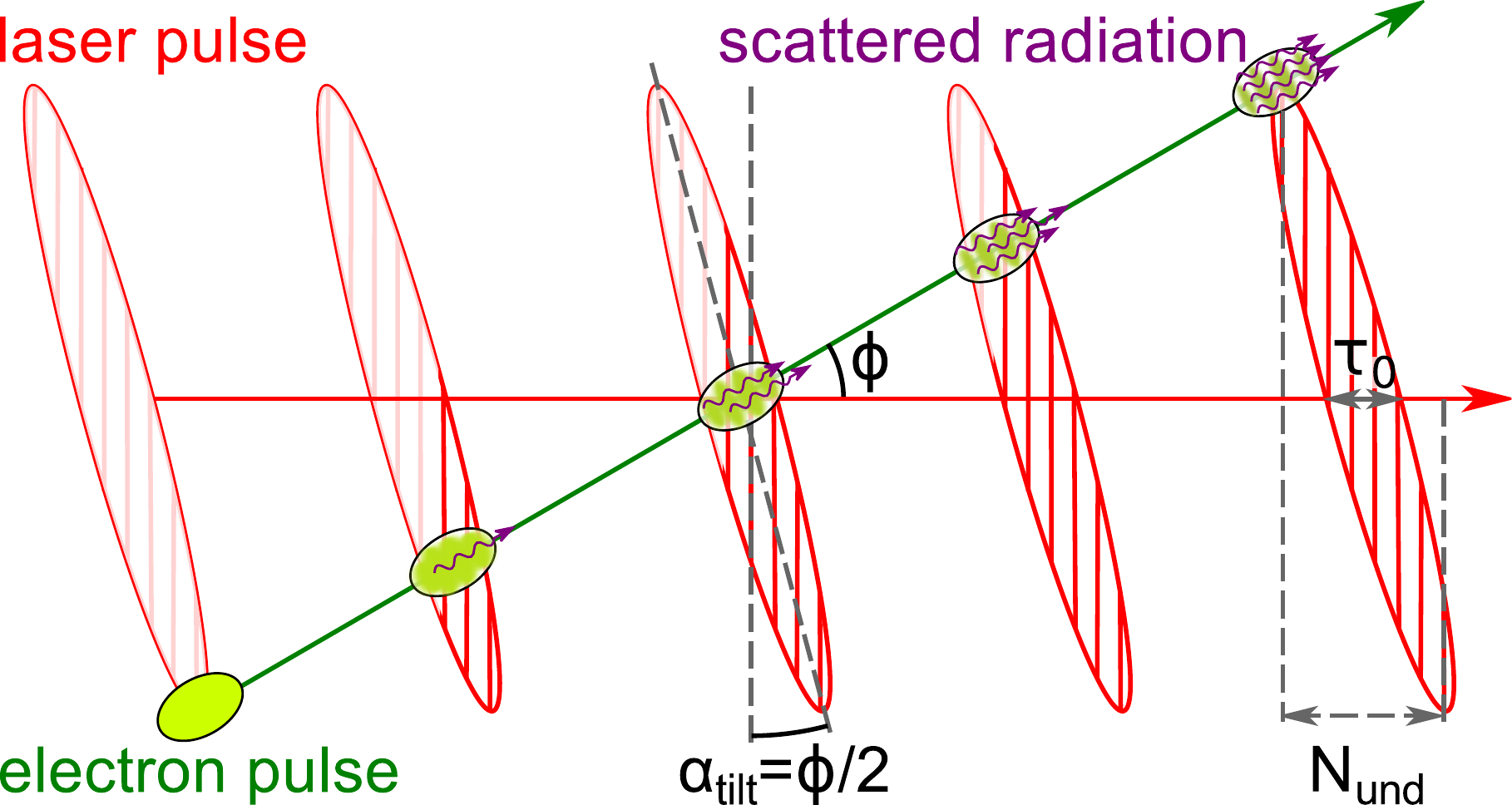}};
    \node[right,align=left] at (0.0,0.0) {(b)};
  \end{tikzpicture}
  \hspace{2cm}
  \begin{tikzpicture}
    \node[anchor=north west,inner sep=0] at (0,0) {\includegraphics[height=4.5cm]{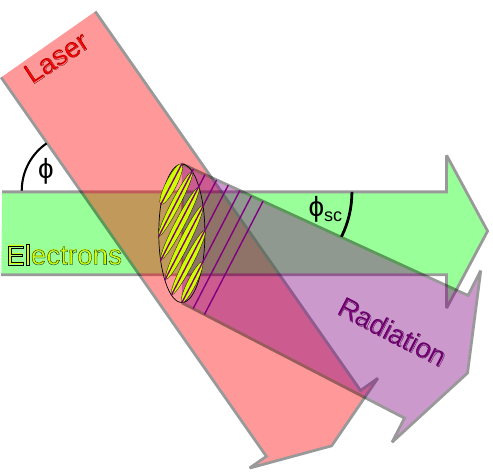}};
    \node[right,align=left] at (0.0,0.0) {(c)};
  \end{tikzpicture}
  \caption{(a) Overview on the Traveling-wave optical FEL experimental design. A laser obtains a pulse-front tilt by passing two gratings and then is focused via a cylindrical mirror, sideways with an incident angle $\phi$ onto an electron beam, where the line focus coincides with the trajectory. Within the interaction zone, (b) shows the TWTS principle of a pulse-front tilted laser pulse subsequently interacting with the focused electron bunch. All parts of the laser get to interact with the electron bunch over an extended interaction length $L_\text{int}$, which can be much longer than the laser Rayleigh length $z_r$, while locally each electron sees a plane wave. This is the basis for both long and narrow optical undulators driving FELs. In terms of FEL physics (c) depicts the difference for the FEL interaction resulting from the sideways incident angle $\phi$ of the TWTS geometry. The resulting FEL radiation features a walk-off angle of $\phi_\text{sc}$ with respect to the propagation direction of the electrons. All graphics shown are derived work based on \cite{Steiniger2014_3}.}
  \label{fig::TWTSOFELIntro}
\end{figure}

A Traveling-wave optical FEL does not collinearly, overlap and collide laser pulse and electron beam in a shared Gaussian-beam type focus.
The Traveling-wave geometry becomes more complex, because it now distinguishes several generally different directions: First the electron direction of electron propagation, secondly the laser direction of propagation, thirdly a laser pulse-front tilt, i.\,e.\ a tilt of the laser pulse envelope with respect to its propagation direction, and finally the FEL radiation direction of propagation.
The laser geometry as shown in Fig.~\ref{fig::TWTSOFELIntro} is a cylindrical focus, where its focal line is collinear with the electron beam.
Therefore the electrons remain in the laser pulse focus over the entire width of the laser pulse which is a big advantage in contrast to head-on scattering geometries.

Consequential the laser direction of propagation is not collinear anymore with the electron beam in the Traveling-wave geometry, but encloses the interaction angle $\phi$ with the electron beam direction of propagation.
Except for angles close to \SI{0}{\degree} and \SI{180}{\degree}, all angles $\phi$ are in principle possible.
Note that the interaction angle $\phi$ changes the FEL resonance condition eq.~\eqref{eq::ThomsonFormular}, which for a given target FEL wavelength requires higher electron energies.
If available laser pulse energy and poor laser-electron beam overlap would impose no technical limits, we would be already done, but also would not require this geometry in the first place. 

The Traveling-Wave Thomson-Scattering (TWTS) geometry therefore makes use of pulse-front tilted, ultra-short laser pulses which allow for full temporal and spatial overlap between electrons and laser pulse, see Fig.~\ref{fig::TWTSOFELIntro}(b).
This enables exploitation of every photon in the laser field for scattering in contrast to head-on geometries.
The laser pulse-front tilt, where the envelope of the laser pulse has an increasing time delay along one transverse direction with respect to the center of the laser pulse, ensures continuous overlap of electrons and laser pulse over the whole laser pulse width.
In this way, the length of the optical undulator is given through the laser beam diameter and only limited by the available laser power as opposed to head-on scattering geometries, where the interaction distance is limited by the laser Rayleigh length.
Thus, almost the total available energy of a high-power laser can be used in TWTS to realize centimeter to meter long optical undulators suitable for OFEL operation in order to produce radiation with higher intensity, narrower bandwidth and consequential higher brilliance than a head-on Thomson scattering geometry.

The required laser pulse-front tilt for a relativistic electron beam is $\alpha_\mathrm{tilt}\simeq\phi/2$. In an experimental setup, these large pulse front tilts can be realized by applying an angular chirp to the laser pulse which can achieved by a pair of optical gratings.
The goal is optimal overlap, in which ideally the entire laser pulse subsequently interacts with the electron bunch.
At a given laser pulse energy, this optimization of overlap vastly improves the coupling between laser and electrons, because the constraints of the Rayleigh limit do not apply anymore for the length of the interaction zone, but merely to the much smaller dimensions of the electron bunch itself.
One main technical challenge is controlling the group delay dispersion and spatial dispersion induced by the angular chirp during propagation.
This can be solved by pre-compensating the dispersion contributions prior to the interaction using two standard optical gratings in addition to the CPA laser systems compressor \cite{Steiniger2018,Steiniger2014_3,Debus2010}.

Utilizing Traveling-Wave Thomson-Scattering does neither change the interaction dynamics of Quantum FELs nor the one of classical FELs on a fundamental level compared to a head-on scattering geometry.
On the level of characteristics differences arise if finite-size electron and laser beams are considered.
In the laboratory frame, as shown in Fig.~\ref{fig::TWTSOFELIntro}(c), the direction of FEL radiation does not exactly coincide with the electron direction of propagation anymore, but features a walk-off angle
\begin{equation}
\phi_\text{sc}=
\begin{cases}
\phi\quad & \text{ , $\phi<\frac{1}{\gamma}$ (forward scattering),}\\
\frac{1}{\gamma_0^2\cdot\phi}\quad & \text{ , $\phi>\frac{1}{\gamma}$ and $\phi\ll 1$, (small-angle back scattering)}\\
\arctan\left(\frac{\sin\phi}{2\gamma^2\cdot(1-\beta_0\cos\phi)}\right)\quad & \text{ , $\phi\gg\frac{1}{\gamma}$ (large-angle back scattering).}
\end{cases}\, ,
\label{eq::TWTSSimpleAngle}
\end{equation}
which is strictly smaller than $1/\gamma$. In many practical cases, this angle $\phi_\text{sc}$ turns out to be negligible. As long as condition
\begin{equation}
  L_\text{g}\cdot\tan\phi_\text{sc} < 2\sqrt{2}\sigma_\text{e}
\label{eq::walkOffConstraint}
\end{equation}
holds, the angle $\phi_\text{sc}$ remains small enough to cause a notable walk-off loss of the FEL laser energy with respect to the electron beam within one gain length \cite{Steiniger2014_3}.

For further details on the Traveling-wave Thomson scattering scheme and possible implementation in high-power lasers we refer the reader to our more extensive works \cite{Steiniger2014_3,Steiniger2014,Steiniger2016,Steiniger2018,Debus2010,Schramm2014,Steiniger2014_2}. While there are many indications, that such TWTS geometries are well in the range of current machines and laboratories, the main practical challenge in experimentally implementing TWTS are system integration efforts, such as for example accompanying diagnostics for the TWTS characteristics, as well as finding suitable space in existing beam lines.

\noindent Summarizing the main advantages for Quantum FELs that possibly can be obtained by the TWTS geometry:

First, the TWTS geometry eliminates the Rayleigh limit of the interaction length. Rayleigh and laser pulse envelope limitations are only imposed on the much smaller electron bunch dimensions. Hence, this saves laser pulse energy, by making it possible to match the laser spot size to specific combinations of electron bunch size and interaction length, without increasing laser spot size and thus laser pulse energy to satisfy the Rayleigh length criterion eq.~\eqref{eq::QFEL_RayleighConstraint}.

Secondly, TWTS reduces intensity variations in optical undulators, since the interaction takes place in the temporal center of the drive laser beam, while the temporal evolution of the optical undulator strength experienced by the electrons is primarily determined by the transverse laser profile. Experimentally, transverse laser profiles are considerably easier to adjust than the temporal profile.

Thirdly, despite local pulse durations being ultrashort, in a TWTS laser, every electron can undergo many optical undulator  periods at uniform field amplitudes as from a temporally extended plane wave. Hence, TWTS strongly reduces the effective bandwidth observed by electrons even for ultrashort lasers.

Finally, a more practical advantage is that tuning the interaction angle $\phi$ gives the freedom to go to different FEL wavelengths at constant electron energy.

\section{Limits to Quantum FELs due to space-charge and spontaneous emission}

\subsection{Space-charge counteracting QFEL dynamics}

Classical FELs theory \cite{Saldin1993,Saldin2000,Geloni2005,Marcus2011} shows that space-charge forces within an electron bunch can negatively impact classical FEL performance and dynamics in several ways. 
From a perspective of electron beam dynamics, space-charge forces change global beam properties \cite{Reiser2008,Hofmann2017}, such as beam size, divergence, energy spread and transverse emittance.
On the microscopic scale however, space-charge forces also act directly against the local density modulations of FEL micro-bunching \cite{Bonifacio1990}, thus altering or inhibiting the FEL dynamics.

Since the focusing fields of magnetic undulators can be used to improve global beam properties, most of theoretical FEL literature investigates the effect of space-charge on micro-bunching.
Hereby, it is useful to introduce the inverse relativistic plasma wavenumber
\begin{equation}
k_\text{p}^{-1}=c\gamma^3/\Omega_\text{p}
\label{eq::relativisticBeamPlasmaFrequency}
\end{equation}
of the electron beam as a characteristic length scale, where $\Omega_\text{p}=\sqrt{e^2 n/\varepsilon_0 m}$ denotes the nonrelativistic plasma frequency. For transversal motions in bunched beams, the strong $\gamma^3$ scaling originates from the transversal relativistic mass $m\gamma$ combined with the reduction of the electric Coulomb repulsion by the magnetic attraction according to the $\vec{v}\times\vec{B}$-term of the Lorentz force $\vec{F_L}=e\vec{E} (1-\beta^2)=e\vec{E}/\gamma^2$. For longitudinal motions there is no such magnetic field reduction, but the longitudinal relativistic mass $\gamma^3 m$ takes the place of the non-relativistic mass~\cite{Reiser2008}.

The parameter $k_\text{p}^{-1}$ describes a characteristic length scale for space-charge evolution. 
For beam propagation distances $L\ll k_\text{p}^{-1}$, space-charge forces become negligible, while for $L\geq k_\text{p}^{-1}$ space-charge forces become relevant for beam dynamics.

For a quantitative understanding of the space-charge dynamics inherent in the FEL process, classical 1D FEL theories often self-consistently introduce space-charge forces as periodic potentials at the micro-bunching period \cite{Saldin2000,Saldin1993,Bonifacio1990}, with its amplitude being derived from Fourier transform components of the longitudinal electron bunch density profile.

As a result one arrives at the prerequisite of a dimensionless space-charge parameter being smaller than unity
\begin{equation}
L_\text{g} \cdot k_\text{p} < 1 \, ,
\label{eq::QFELSpaceChargeRaman}
\end{equation}
which amounts to the statement that micro-bunching of the electron beam has to grow faster than the space-charge potential-equalizing dynamics.
If $L_\text{g}\cdot k_b$ reaches an appreciable fraction of unity, this leads to increased gain lengths, as well as to shifted FEL wavelengths, which marks the beginning of the so called Raman regime.
The FEL regime free of noticeable space-charge dynamics is called Compton regime in standard FEL literature, see e.\,g.\ ref.\ \cite{Bonifacio1990}.
In the following we adopt this distinction of Raman and Compton regimes for the Quantum FEL. 

Despite first attempts by Serbeto~\emph{et~al.} using a semi-classical quantum fluid approach \cite{Serbeto2008,Serbeto2009,Monteiro2013}, there exists no comprehensive theory yet on the 6D phase-space properties and dynamics of QFELs under the influence of space-charge.
A theory of the Raman QFEL regime would have to take the maximum allowable relative momentum spread into account, cf.\ eq.~\eqref{eq:ladder_two}, before competing momentum transitions occur.
These processes effectively disrupt the isolated two-level quantum system.

Accordingly, it is useful to define an additional, more defensive requirement, which excludes significant space-charge effects not only over one gain length, but over the entire QFEL interaction length
\begin{align}
L_\text{int} \cdot k_\text{p} &< 1 \nonumber\\
N_\mathrm{g} L_\text{g} \cdot k_\text{p} &< 1\,,
\label{eq::QFELSpaceChargeQFEL}
\end{align}
where $N_\text{g}$ denotes the number of gain lengths $L_\text{g}$ within the interaction length $L_\text{int}$.
In analogy to classical FEL theory we assume that the earliest saturation of a Quantum SASE FEL occurrs at approximately $N_\text{g,sat}\approx 10$.
While on the one hand the limit in eq.~\eqref{eq::QFELSpaceChargeQFEL}, corresponding to virtually all QFEL theories published so far, could be too strict with regard to potentially realizable two-level QFELs in the Raman regime, the Raman regime condition eq.~\eqref{eq::QFELSpaceChargeRaman} on the other hand will be too permissive, since it includes neither increasing gain lengths nor the break-down of the two-level quantum system due to space-charge.
Thus we discuss space-charge in both these limits in the following -- keeping in mind that the true space-charge constraint likely lies in between these two extremes.

It is useful to restate eqs.~\eqref{eq::QFELSpaceChargeRaman} and \eqref{eq::QFELSpaceChargeQFEL} to better expose all fundamental dependencies.
For this purpose we use the relation $L_\text{g}=\lambda_u/4\pi\Gamma$ with $\Gamma=\rho \sqrt{\bar{\rho}/(1+\bar{\rho})}$ and isolate the direct dependency with respect to $\bar{\rho}$ for $a_0\ll 1$ by applying the definition eq.~\eqref{eq::QFELparameter}. Then we transform $k_\text{p}$ using eqs.~\eqref{eq::relativisticBeamPlasmaFrequency} and \eqref{eq::defineRho} in order to cancel the lowest order $\rho^{3/2}$-dependency of $L_\text{g}$ and arrive at
\begin{equation}
L_\text{int} k_\text{p}
    = N_\text{g} \sqrt{\frac{8 \lambda_c}{\gamma\lambda_\text{FEL}} \frac{(1+\bar{\rho})}{a_0^2}} \leq 1\,\text{, with }
    \begin{cases}
        N_\text{g} = (L_\text{int}/L_\text{g})\, , \text{for QFELs with negligible space-charge \emph{(Compton regime)}}\, ,\\
        N_\text{g} = 1\, , \text{for QFELs with significant space-charge \emph{(Raman regime)}}\, .
    \end{cases}\, .
\label{eq::QFEL_QFELRamanCond}
\end{equation}

Since the electron-light interaction has to occur faster than the space-charge dynamics, solving eq.~\eqref{eq::QFEL_QFELRamanCond} for $a_0$ yields a minimum requirement for the normalized laser amplitude
\begin{align}
a_\text{0} &\geq a_\text{0,min}\,\text{, with}\\
a_\text{0,min} &= \sqrt{\frac{8 \lambda_c}{\gamma\lambda_\text{FEL}} (1+\bar{\rho})} N_\text{g}
\label{eq::QFEL_Min_a0}
\end{align}
where again $N_\mathrm{g} = L_\text{int}/L_\text{g}$ and $N_\text{g} = 1$ for the Compton QFEL regime and the Raman QFEL regime, respectively.

\subsection{Spontaneous emission counteracting QFEL coherence}

Another major process competing with the QFEL stimulated emission dynamics is the spontaneous decay of electron momentum in the laser undulator field by standard Compton scattering.
Spontaneous emission is polychromatic in nature, since it is robustly a 3D effect, where the emitted wavelength depends on the angle of emission $\theta$, cf.\ eq.~\eqref{eq::ThomsonFormular}, and ranges from $\lambda_u$ to $\lambda_\text{FEL}$ in its lowest order.

Thus the great majority of spontaneously emitted photons lead to electrons decaying to momentum states outside the momentum bands allowed by the QFEL, see also \cite{Robb2013} and corresponding comment.
If all electrons decayed into other momentum states, the QFEL dynamics would not take place.
Hence we require that the number of coherent photons emitted by a Quantum FEL surpasses the number from incoherent Compton scattering. In other words, all QFEL physics has to take place within the lifetime of the initial electron momentum states.

For investigating the influence of spontaneous emission, the two-mode model of the Hamiltonian in eq.~\eqref{eq:qtheory_H} is insufficient. Instead, one has to consider an electron, which initially is prepared in the excited state $p=q/2$ of the average rest frame, coupled to a reservoir of all possible modes of the electromagnetic field with each mode being in the vacuum state $\ket{0}$ except for the undulator mode.

In a standard Wigner-Weisskopf approach~\cite{scullylamb} one can derive the decay constant~\cite{rainer}
\ba\label{eq:qtheory_Gamma}
D = 4\pi^2\underbrace{\frac{V}{(2\pi^3)}\frac{\omega^2}{c^3}}_{\text{all modes}}\underbrace{\frac{8\pi}{3}}_{\text{3D}}\underbrace{g^2}_{\text{interaction}}
\ea
which denotes the inverse life time of the excited state in the average rest frame before it incoherently decays due to spontaneous emission.
The origin of the single terms in eq.~\eqref{eq:qtheory_Gamma} is as follows: the first contribution emerges due to the conversion of a sum over all modes to an integral by means of the density of states.
The second term $8\pi/3$ represents a three-dimensional effect and arises by integrating over all possible directions of polarization.
Moreover, we obtain a quadratic dependency on the coupling $g$ of electron and fields which is analogous to Fermi's golden rule~\cite{fedorov}.
We emphasize that in the derivation of eq.~\eqref{eq:qtheory_Gamma} collective effects~\cite{dicke} due to the simultaneous interaction of many electrons with the fields were neglected. 
We further assumed $a_0\ll 1$ and neglected multi-photon processes, as well as cases where the magnitude of the recoil becomes similar to the particle momentum at high energies.
If the latter assumptions do not hold, the constraint due to spontaneous emission becomes even more restrictive.

The classical result~\cite{Jackson1998} for the inverse decay length $R_\text{sp}^{-1}$ of a single electron in the laboratory frame reads
\ba
\label{eq:qtheory_R}
R_\text{sp}^{-1}=\frac{2\pi\alpha_f}{3}\frac{a_0^2}{\lambda_u}\, ,
\ea
which is also the expression used in Ref.~\cite{Robb2011} with $\alpha_f$ denoting the fine structure constant. 
The above inverse decay length $R_\text{sp}^{-1}\equiv D/c$ is consistent with the result \eqref{eq:qtheory_Gamma} from the quantum theory~\cite{rainer}.

In order to obtain FEL amplification the lifetime of the excited state has to exceed the typical interaction time, that is, the electron has to have enough time to interact coherently with the undulator and the FEL mode before it spontaneously emits radiation.
Hence, we require the inverse decay length $R_\text{sp}^{-1}$ to be smaller than half the total QFEL interaction length
\begin{align}
	R_\text{sp}^{-1} \cdot L_\text{int} &\leq \frac{1}{2}
\end{align}
or
\begin{align}
	\frac{\alpha_f}{6\sqrt{3}\lambda_c}N_g a_0^2\gamma\lambda_\text{FEL}\sqrt{\frac{1+\bar{\rho}}{\bar{\rho}^3}} &\leq \frac{1}{2} \, .
	\label{eq::spontaneousEmission}
\end{align}
Note, that a higher drive laser intensity leads to more spontaneous emission $R_\text{sp}^{-1}\propto a_0^2$.
Accordingly, this results in the constraint
\begin{align}
	a_0 &\leq  a_\text{0,max}\,\text{, with} \\
	a_\text{0,max} &= \sqrt{ \frac{3\lambda_c}{\alpha_f\gamma\lambda_\text{FEL}N_\text{g}}\sqrt{\frac{\bar{\rho}^3}{1+\bar{\rho}}}}
\label{eq::QFEL_max_a0}
\end{align}
where $N_\mathrm{g} = L_\mathrm{int} / L_\mathrm{g}$ is independent of the regime.
This limit is opposed to the space-charge requirement \eqref{eq::QFEL_Min_a0} of a minimum $a_0$ and features the same dependency on $\lambda_\text{FEL}$ and $\gamma$.

\subsection{Strong limit by opposed scalings of constraints}

The opposing constraints of eq.~\eqref{eq::QFEL_Min_a0} and \eqref{eq::QFEL_max_a0} from space-charge and spontaneous emissions now impose strong limitations to the feasibility of any Quantum FEL.

There are four fundamental quantities: the quantum parameter $\alpha\equiv\bar{\rho}^{3/2}$, the undulator length $N_\text{g}$ in units of the gain length, as well as the number of spontaneous decays $R_\text{sp}^{-1}\cdot(L_\text{g} N_\text{g})$ at the end of the interaction and beam plasma wavenumber times the undulator length $k_\text{p} L_\text{g} N_\text{g}$.
We now express the undulator length $N_\text{g}$ as a function of the other three quantities.
In addition we have to simultaneously meet three constraints: a small quantum parameter $\alpha=\bar{\rho}^{3/2}\ll 1$ as well as the limits set by the number of spontaneous decays and space-charge effects to Quantum FEL operation, eq.~\eqref{eq::QFEL_max_a0} and eq.~\eqref{eq::QFEL_Min_a0} respectively.

\begin{figure}[!t]
  \centering
  \begin{tikzpicture}
    \node[anchor=north west,inner sep=0] at (0,0) {\includegraphics[height=5.5cm]{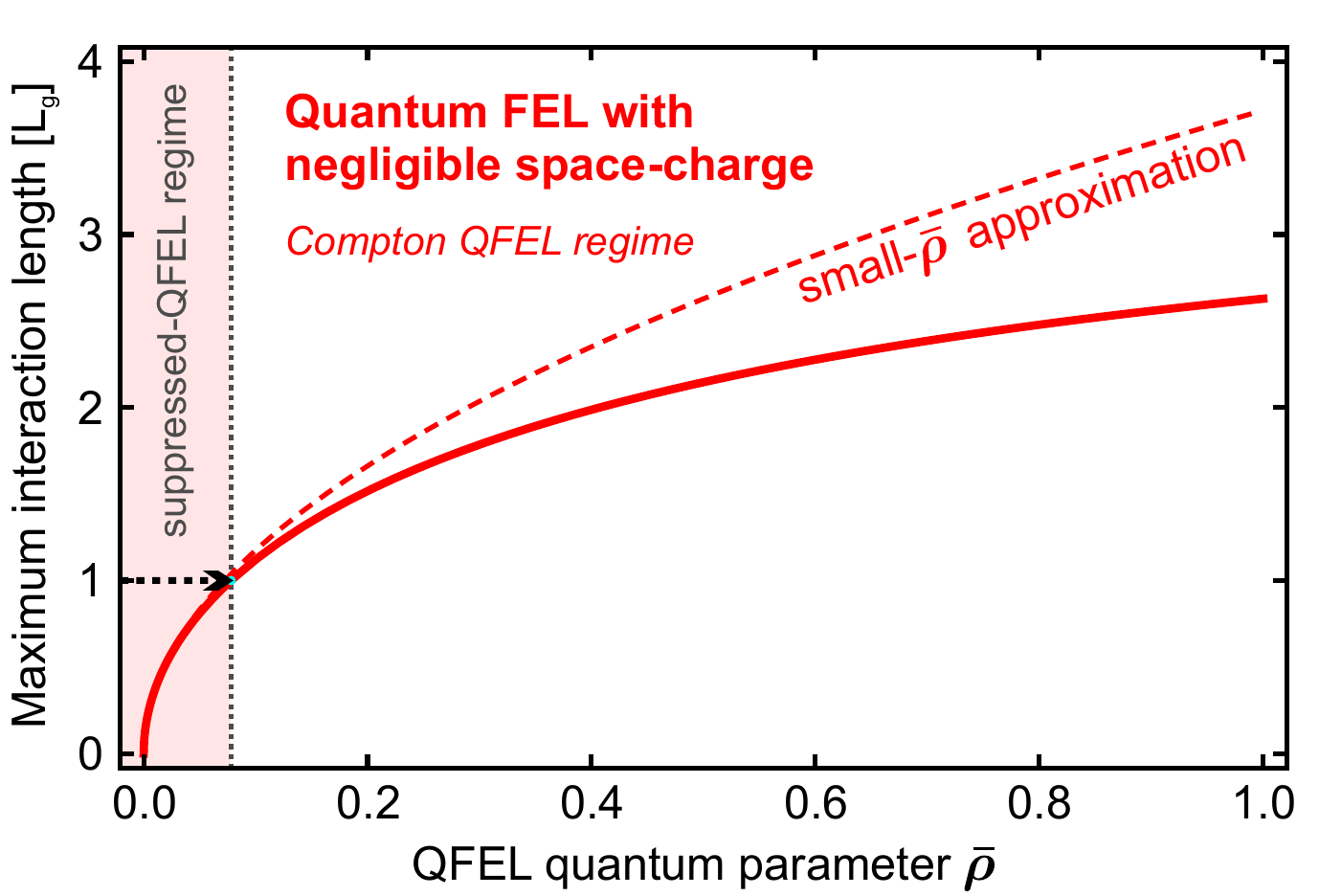}};
    \node[right,align=left] at (0.0,0.0) {(a)};
  \end{tikzpicture}
  \hfill
  \begin{tikzpicture}
    \node[anchor=north west,inner sep=0] at (0,0) {\includegraphics[height=5.5cm]{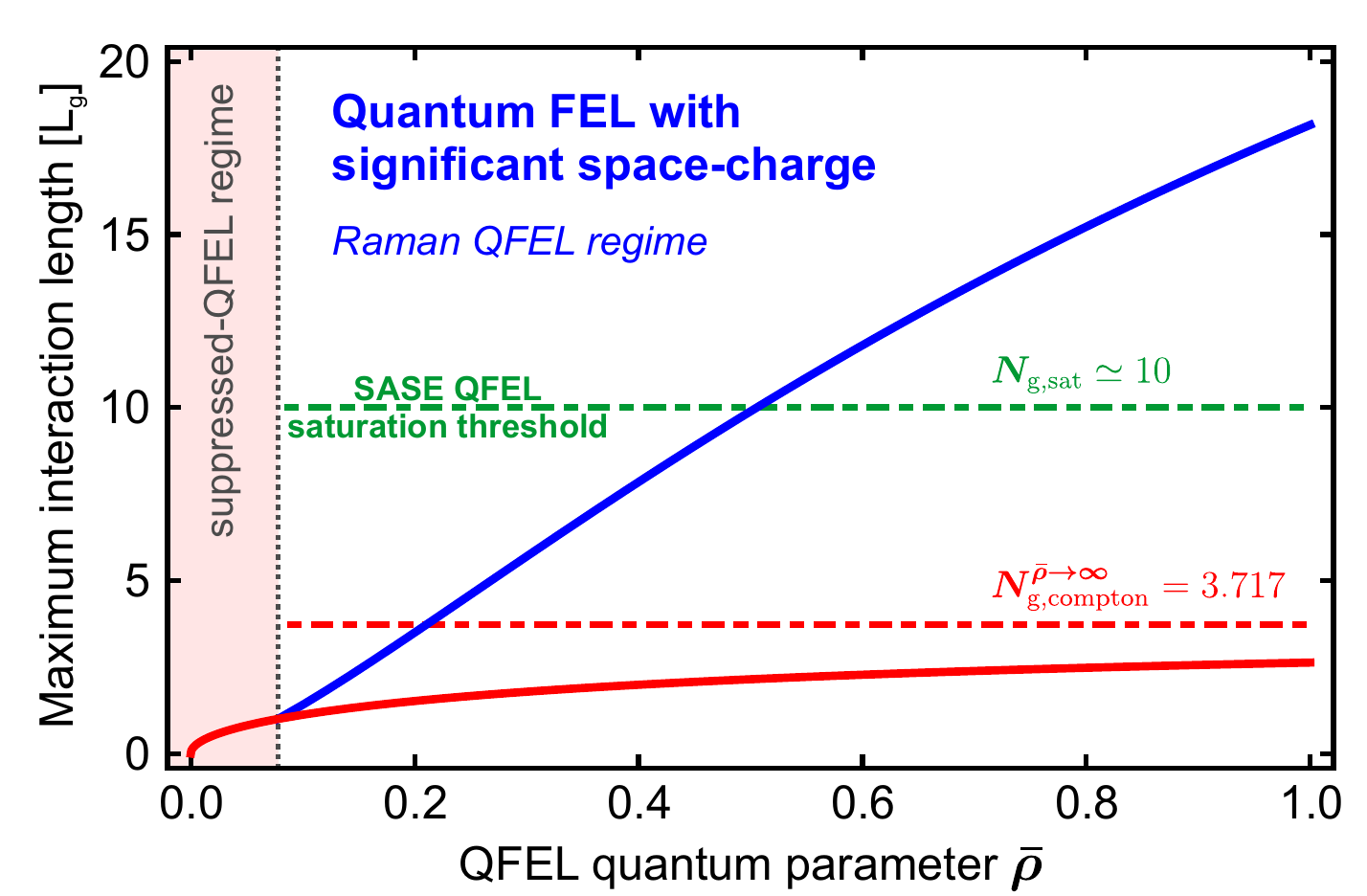}};
    \node[right,align=left] at (0.0,0.0) {(b)};
  \end{tikzpicture}
  \caption{Depicts the maximum number of achievable gain lengths $N_\text{g}$ for Quantum FELs with respect to the quantum parameter $\bar{\rho}$ based on the combined limitations due to space-charge and spontaneous emission.
  We distinguish between two models, representing lower and upper constraints of a future, more rigorous QFEL theory taking space-charge, as well as spontaneous emissioninto account:
  (a) illustrates the \emph{Compton QFEL regime} according to eq.~\eqref{eq::SCSP_QFELLimit}, which assumes negligible space-charge dynamics $L_\text{int}\cdot k_p<1$.
  The dashed curve shows the small-$\bar{\rho}$ approximation of eq.~\eqref{eq::SCSP_SimpleQFELLimit}.
  (b) depicts eq.~\eqref{eq::SCSP_RamanLimit} of the supposed \emph{Raman QFEL regime} in which, akin to classical FELs, space-charge dynamics is an integral part of the FEL interaction $L_\text{g}\cdot k_p<1$.
  The red-dashed curve shows the Compton regime for comparison.
  The green-dashed line marks the undulator length at which SASE operation would become possible.}
  \label{fig::QFEL_SpaceCharge_SpontEmission_ScalingGraphs}
\end{figure}

Searching for the largest acceptable undulator length $N_\text{g,max}$ for $a_\text{0,max} \geq a_\text{0,min}$ and $\alpha\ll 1$ we obtain the simple relation
\begin{equation}
N_\text{g,max}=\left(\frac{3}{8}\frac{\alpha}{\alpha_f} \right)^{1/3}\, ,
\label{eq::SCSP_SimpleQFELLimit}
\end{equation}
which in particular is independent of $\lambda_\text{FEL}$ and $\gamma$.
This dimensionless undulator length $N_\text{g,max}$ now can be compared, see dashed line in Fig.~\ref{fig::QFEL_SpaceCharge_SpontEmission_ScalingGraphs}, with a saturation length $N_\text{sat}$, showing that the largest acceptable undulator length $N_\text{g,max}=3\,$-$\, 4$ is much smaller than the typical saturation length $N_\text{sat}\approx 10$ for SASE FELs.
Therefore the maximum photon yield is diminished by a factor of $\e{N_\text{g,max}-N_\text{sat}}$.

After presenting the core argument in a nutshell, we reiterate above result of eq.~\eqref{eq::SCSP_SimpleQFELLimit} in a more general form without the condition $\alpha=\bar{\rho}^{3/2}\ll 1$ while adopting the more commonly used notation $\bar{\rho}$ for the quantum parameter.
The maximum undulator length is for example obtained by reformulating the spontaneous emission criterion \eqref{eq::spontaneousEmission} to a statement on $N_\mathrm{g}$.
Then the largest possible $N_\mathrm{g}$ is found by inserting $a_{0,\mathrm{min}}$ for $a_0$.

In the Compton QFEL regime where we require negligible space-charge effects on the interaction dynamics the largest possible undulator length is
\begin{equation}
N_\text{g,max}^\text{Compton} = \left( \frac{3}{8 \alpha_f} \sqrt{\frac{\bar{\rho}^3}{(1 + \bar{\rho})^3}} \right)^{1/3}
\label{eq::SCSP_QFELLimit}
\end{equation}
where we have used the defensive space-charge criterion eq.~\eqref{eq::QFELSpaceChargeQFEL}.
Figure~\ref{fig::QFEL_SpaceCharge_SpontEmission_ScalingGraphs}(a) shows the scaling of $N_\text{g,max}^\text{Compton}$ as a function of the quantum parameter $\bar\rho$.

In contrast to the Compton QFEL regime, the more relaxed result
\begin{equation}
N_\text{g,max}^\text{Raman} = \frac{3}{8 \alpha_f} \sqrt{\frac{\bar{\rho}^3}{(1 + \bar{\rho})^3}}
\label{eq::SCSP_RamanLimit}
\end{equation}
is based upon the optimistic assumption of a hypothetic Raman QFEL regime, in which space-charge forces alter the Quantum FEL dynamics but without destroying either the two-level system nor the QFEL gain.
The $N_\text{g,max}-\bar{\rho}$-scaling following eq.~\eqref{eq::QFELSpaceChargeRaman} is depicted in Fig.~\ref{fig::QFEL_SpaceCharge_SpontEmission_ScalingGraphs}(b).

The defensive criterion in eq.~\eqref{eq::SCSP_QFELLimit} on the one hand suggests that it is impossible to operate any Quantum FEL $\bar{\rho}\leq 0.5$ in a regime where space-charge can be neglected for more than 3 to 4 gain lengths $L_\text{g}$.
Especially, there is no ``safe'' regime for a Quantum SASE FELs in which both spontaneous emission and space-charge dynamics are negligible for the extent of a full QFEL saturation length. 
The limit
\begin{equation}
\lim\limits_{\bar{\rho}\to\infty}N_\text{g}^\text{Compton}=3.71782\, ,
\end{equation}
shows that classical SASE FEL dynamics at saturation is always subject to significant space-charge or spontaneous emission dynamics or both.

The probably overoptimistic criterion in eq.~\eqref{eq::SCSP_RamanLimit} on the other hand shows that even in a permissive quantum Raman regime -- if it exists -- the deep quantum regime $\bar{\rho}\ll 1$ remains inaccessible since the stimulated QFEL interaction can last only for very few gain lengths before the dynamics is dominated by spontaneous emission or space-charge effects.
According to the optimistic scaling of Fig.~\ref{fig::QFEL_SpaceCharge_SpontEmission_ScalingGraphs}(b), a Quantum SASE FEL only becomes possible for $\bar{\rho}\geq 0.5$.
Both scalings show that within the deep quantum regime, a QFEL necessarily needs to enter the Raman QFEL regime, otherwise there would be too many spontaneous emission events.
However, according to the fitting formula for the classical gain length in \emph{Marcus~et.~al.}~\cite{Marcus2011}
\begin{equation}
\tilde{L}_{g,sc} \geq \tilde{L}_\text{g} ( 1+(2 k_\text{p} \tilde{L}_\text{g})^{1.91})\,\text{, with }\tilde{L}_\text{g}=L_\text{g}/\sqrt{3}
\end{equation}
it increases with $2 k_\text{p} \tilde{L}_\text{g}$.
This suggests that the deep quantum regime $\bar{\rho}<0.078$ following from $N_\text{g}^\text{Compton}(\bar{\rho})<1$ is robustly forbidden due to excessive spontaneous emission events indirectly caused by space-charge.

First, above estimates rely on the known micro-bunching dynamics of classical FEL and the classical energy loss due to spontaneous emission.
In addition to a quantum treatment, more detailed studies would need to take the continuous momentum distribution of spontaneously emitted photons into account, as well as the statistics of spontaneous scattering events instead of the mean scattering probability.
Although these contributions can substantially add to the numerical complexity of the analytical expressions, we assume that the basic scaling with $\bar{\rho}$ does not change.

Secondly, the space-charge dynamics of the global electron bunch evolution, in contrast to micro-bunching, is not directly taken into account. Although the space-charge parameter with the beam plasma frequency is related to the generalized perveance in beam dynamics \cite{Reiser2008}, hence providing a useful time-scale for the space-charge dynamics, its results do not provide quantitative estimates on beam parameters. As shown in a later example, quantitative space-charge calculations of realistic beams usually cannot be modeled analytically anymore, but require numerical particle-tracking simulations. Since the space-charge dominated regime cannot be avoided for Quantum FELs, it is this dynamics which is most destructive to the possibility of experimentally realizing Quantum SASE FELs.

\section{Designing Quantum FEL experiments}

\begin{figure}[!t]
  \centering
  \begin{tikzpicture}
    \node[anchor=north west,inner sep=0] at (0,0) {\includegraphics[height=8.5cm]{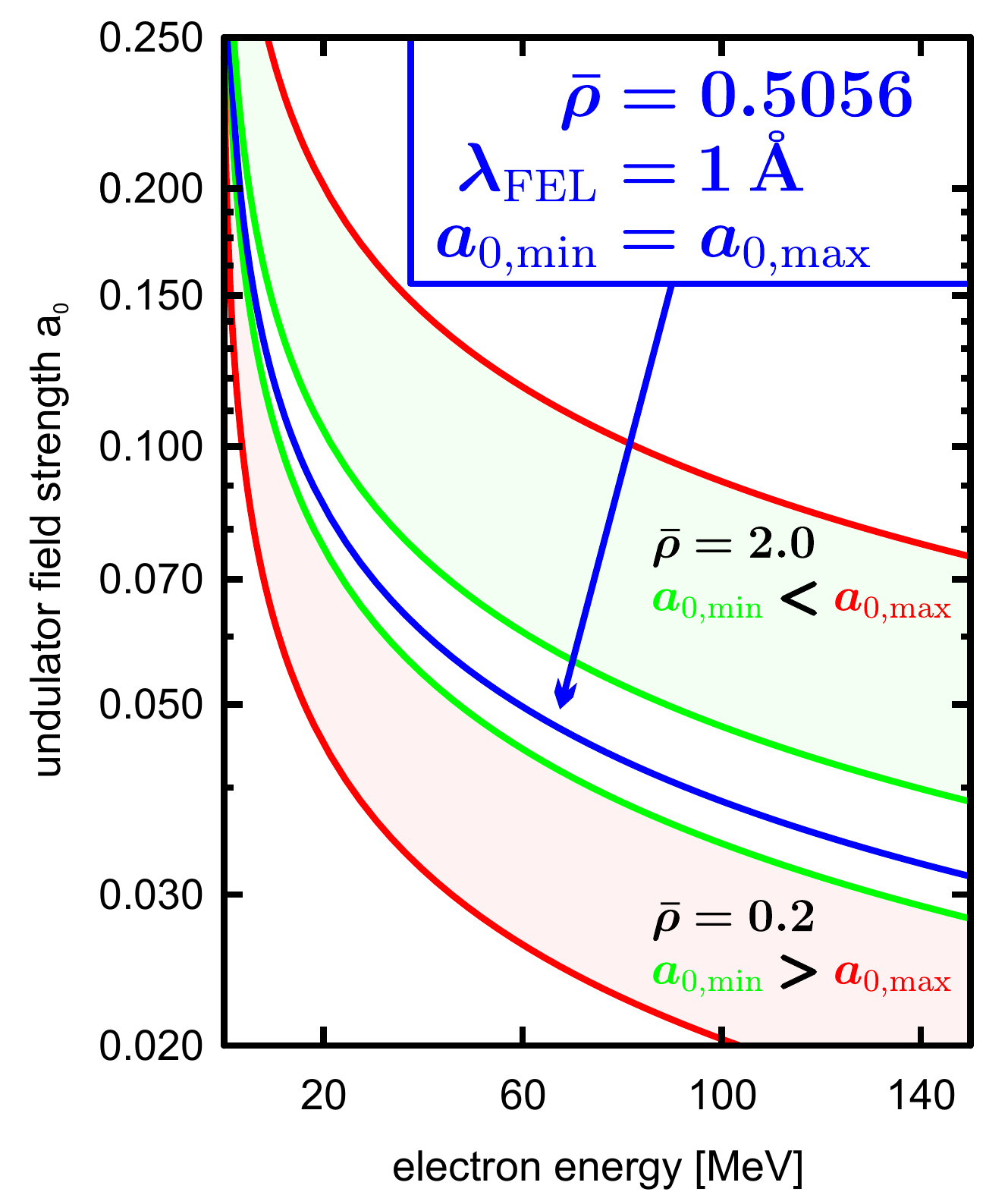}};
    \node[right,align=left] at (0.0,0.0) {(a)};
  \end{tikzpicture}
  \hspace{1cm}
  \begin{tikzpicture}
    \node[anchor=north west,inner sep=0] at (0,0) {\includegraphics[height=8.5cm]{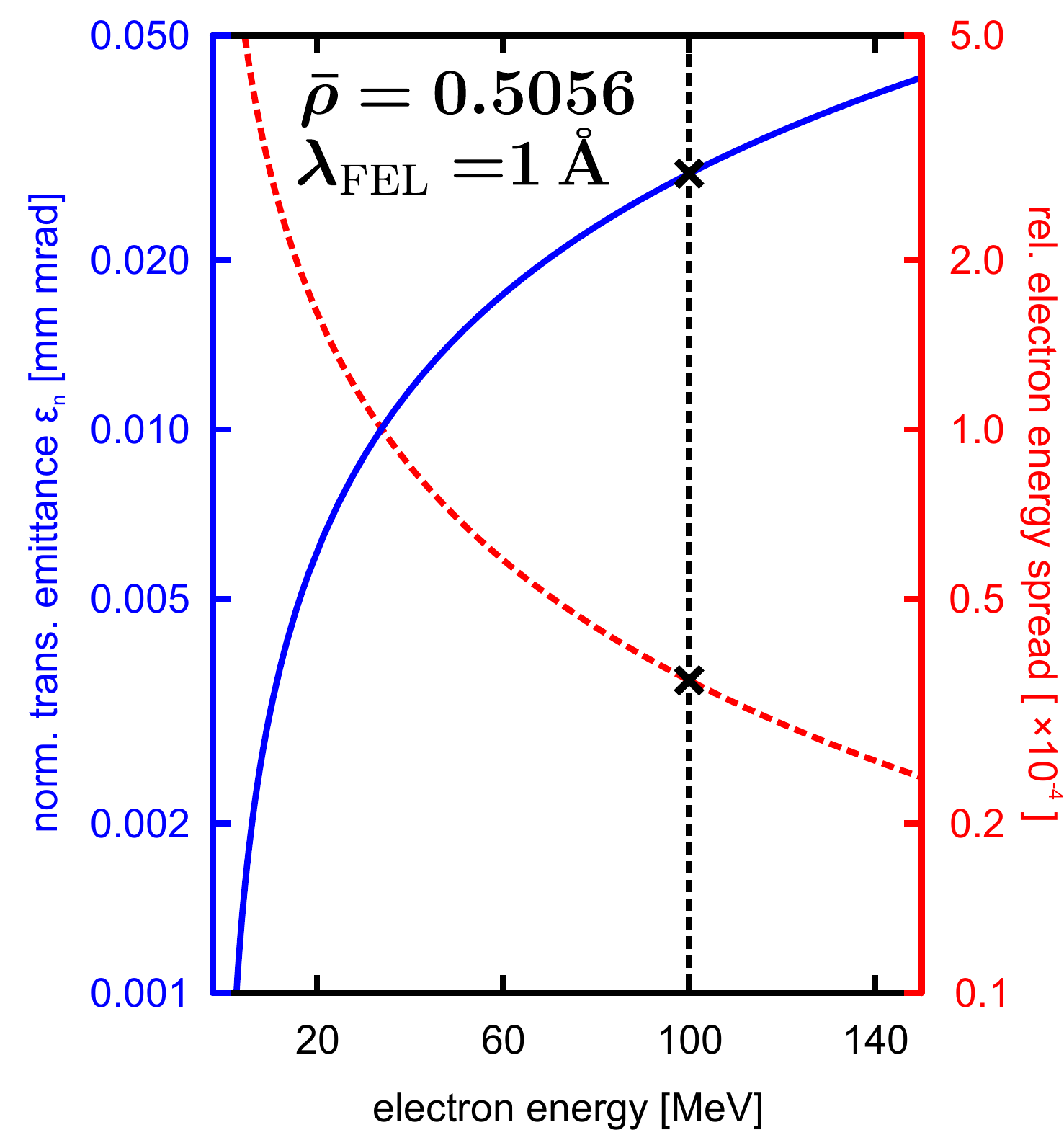}};
    \node[right,align=left] at (0.0,0.0) {(b)};
  \end{tikzpicture}
  \caption{Illustrates the major decisions for a Quantum SASE FEL design satisfying $\lambda_\text{FEL}=\SI{1}{\angstrom}$, $\bar{\rho}=0.5056$ and $\hat{\varepsilon}=10.0$.
  (a) shows the minimum and maximum allowed laser strength $a_0$, see eqs.~\eqref{eq::QFEL_Min_a0} and \eqref{eq::QFEL_max_a0}, for several values of $\bar{\rho}$ according to the constraints on space-charge and spontaneous emission in the optimistic case of a Raman QFEL regime.
  While the value $\bar{\rho}=2.0$, which tends to a classical FEL regime, allows for a wide range of possible $a_\text{0,min}<a_\text{0,max}$ (green) for any given electron energy, the deep quantum regime at $\bar{\rho}=0.2$ remains excluded since $a_\text{0,min} > a_\text{0,max}$ (red).
  The blue curve presents the limit of $a_\text{0,min} = a_\text{0,max}$, thus marking the lowest permissible quantum parameter $\bar{\rho}=\num{0.5056}$ for a Quantum SASE FEL.
  (b) depicts the choice of the target electron energy with its consequences on the energy spread and transverse emittance requirements.
  The dashed line marks the target electron energy chosen for the design in tab.~\ref{tab::QFELDesignExample}.}
  \label{fig::QFEL_DesignChoices}
\end{figure}

For Quantum SASE FELs there are essentially only four main quantities that have the largest impact on selecting a suitable experimental design:

First, one needs to pick a desired FEL wavelength and a target quantum parameter $\bar{\rho}$.
In our example we aim for an \AA-wavelength QFEL.
Since we try to enter the quantum regime ($\bar{\rho}\ll 1$), but know about the limits on the achievable interaction length arising from spontaneous emission and space-charge, we pick the quantum parameter by choosing the lowest value $\bar{\rho}=0.5056$ which allows to keep the interaction going over at least ten gain lengths.
The range of possible values for $\bar{\rho}$ fulfilling this requirement can be seen in Fig.~\ref{fig::QFEL_DesignChoices}(a) showing the scaling of the space-charge and spontaneous emission constraints on $a_0$, eqs.\ \eqref{eq::QFEL_Min_a0} and \eqref{eq::QFEL_max_a0} respectively, in dependence of the electron energy.
While for values $\bar{\rho}<0.5056$ these constraints on $a_0$ exclude each other ($a_{0,\mathrm{min}} > a_{0,\mathrm{max}}$), there is a range of possible $a_0$ for values $\bar{\rho}>0.5056$ which ensure an interaction length of at least ten gain length in a Raman QFEL regime.

The next decision is on a desired electron energy according to eq.~\eqref{eq::ThomsonFormular}.
It is only due to the TWTS interaction geometry and its variability with respect to the interaction angle, that this choice exists.
According to eqs.\ \eqref{eq::energySpreadConstraint} and \eqref{eq::QFEL_tranverseCoherence} this choice allows to define the sweet spot between the contrary energy spread and transverse emittance requirements for some electron source.
Figure \ref{fig::QFEL_DesignChoices}(b) illustrates the scaling of these requirements in dependence of the electron energy.
The emittance graph is based on a weak requirement on the transverse coherence parameter of $\hat{\varepsilon}=10.0$.
At \SI{100}{MeV} we find the largest transverse emittance goal $\varepsilon_n=\SI{0.031}{mm.mrad}$ with a relative energy spread barely beyond $10^{-5}$.

Then one needs to choose some normalized laser strength $a_0$ for the undulator field from the range $[a_\text{0,min},a_\text{0,max}]$.
If electrons and optical undulator shall interact over the largest possible undulator length of 10\,$L_\mathrm{g}$, being determined by the picked value of $\bar\rho=0.5056$, the maximum and minimum value of $a_0$ are equal. 
In our example $a_0=\num{3.85e-2}$ according to eq.~\eqref{eq::QFEL_Min_a0} or \eqref{eq::QFEL_max_a0}.

\begin{table}[!t]
    \centering
    \caption{Example of a hypothetic \SI{1}{\angstrom} Quantum SASE FEL design.
    At this point the space-charge evolution of the global beam dynamics, see Fig.~\ref{fig::QFEL_GPTsims}, has not been taken into account yet.
    The design is based on the optimistic scaling with regard to space-charge and spontaneous emission constraints as shown in Fig.~\ref{fig::QFEL_SpaceCharge_SpontEmission_ScalingGraphs}(b) (blue curve).}
    \label{tab::QFELDesignExample}
    \scalebox{1.0}{
    \begin{threeparttable}
    \begin{tabular}{>{\kern-\tabcolsep}lc<{\kern-\tabcolsep}}
    \toprule
    \multicolumn{2}{l}{\textbf{Electron bunch}}\\
    \midrule
    electron energy & \SI{100}{MeV}\\
    norm. transv. emittance $\varepsilon_n$ & \SI{0.031}{mm.mrad}\\
    peak current $I_\text{e}$ & \SI{2.0}{kA}\\
    bunch radius (rms) $\sigma_\text{e}$ & \SI{3.0}{\um}\\
    \midrule
    \multicolumn{2}{l}{\textbf{Laser}}\\
    \midrule
    laser wavelength & \SI{1.035}{\um}\\
    TWTS interaction angle $\phi$ & \SI{30.0}{\degree}\\
    laser strength $a_0$ & 0.039\\
    laser intensity & \SI{1.89e+15}{W/\cm^2}\\
    \midrule
    \multicolumn{2}{l}{\textbf{Quantum FEL}}\\
    \midrule
    FEL radiation wavelength & \SI{0.1}{nm}\\
    undulator wavelength $\lambda_u$ & \SI{7.74}{\um}\\
    radiation walk-off angle $\phi_\text{sc}$ & \SI{0.048}{mrad}\\
    Pierce parameter $\rho$ & \num{6.236e-05}\\
    QFEL parameter $\bar{\rho}$ & 0.5056\\
    rel.\ QFEL bandwidth $\Gamma$ & \num{3.61e-05}\\
    1D QFEL gain length $L_\text{g}$ & \SI{17.0}{mm}\\
    maximum interaction distance\tnote{a} & \SI{0.113}{m}\\
    assumed interaction distance $L_\text{int}$ & $\SI{0.170}{m} = \num{22.0e+3} \lambda_u = 10.0 L_\text{g}$\\
    \midrule
    \multicolumn{2}{l}{\textbf{Requirements for }$\bm{L_\text{int}}$}\\
    \midrule
    rel. energy spread & \num{3.61e-05}\\
    laser power & \SI{12.1}{TW}\\
    laser irradiance variation (wavel. shift) & \SI{9.7}{\percent}\\
    laser irradiance variation (pond. deflection) &\SI{63.0}{\percent}\\
    \midrule
    \multicolumn{2}{l}{\textbf{Regime indicators}}\\
    \midrule
    X-ray beam diffraction $<1$? & 0.01 \\
    X-ray radiation walk-off $<1$? & 0.06 \\
    Trans. coherence parameter $\hat{\varepsilon}\in [0.5\ldots 10.0]$? & 10.0\\
    Space-charge parameter $k_\text{p}^{-1} L_\text{g}<1$? & 1.0 \\
    Avg. $\#$ of spontaneous emission events per electron $<1$ & 0.50\\
    Cooperation length $L_\text{c}$ & \SI{2.2}{\um} \\
    \bottomrule
    \end{tabular}
    \begin{tablenotes}
    \item [a] Beyond this distance loss of laser-electron beam overlap, which can be compensated by more laser power through larger laser beam diameter.
    \end{tablenotes}
    \end{threeparttable}
    }
\end{table}%

At this point there are already enough conditions to fix the rest of the parameters.
Remaining electron and laser pulse parameters, such as bunch length and charge as well as wavelength, incidence angle and duration respectively, are constrained by the electron bunch peak current $I_\text{e}$, laser strength $a_0$ as well as the properties of the beam overlap.
An overview on the parameters of this example is given in Table~\ref{tab::QFELDesignExample}.

We emphasize that this example already assumes the existence of a Raman QFEL regime.
Further note that the laser requirements are relatively benign due to the low value of $a_0$.
Thus basic feasibility primarily is determined by the technical feasibility of the electron source and the space-charge dynamics.

As a way to reduce the technical challenges for such electron sources it was suggested in Ref.\ \cite{Bonifacio2017} to aim for low-charge electron bunches with high peak currents. While the minimum bunch radius is given by the transverse emittance criteria for the electron beam, the longitudinal bunch length can in principle be chosen as short as a cooperation length $L_\text{c}$ before the QFEL amplification dynamic changes.

\begin{figure}[!t]
  \centering
  \begin{tikzpicture}
    \node[anchor=north west,inner sep=0] at (0,0) {\includegraphics[width=0.99\textwidth]{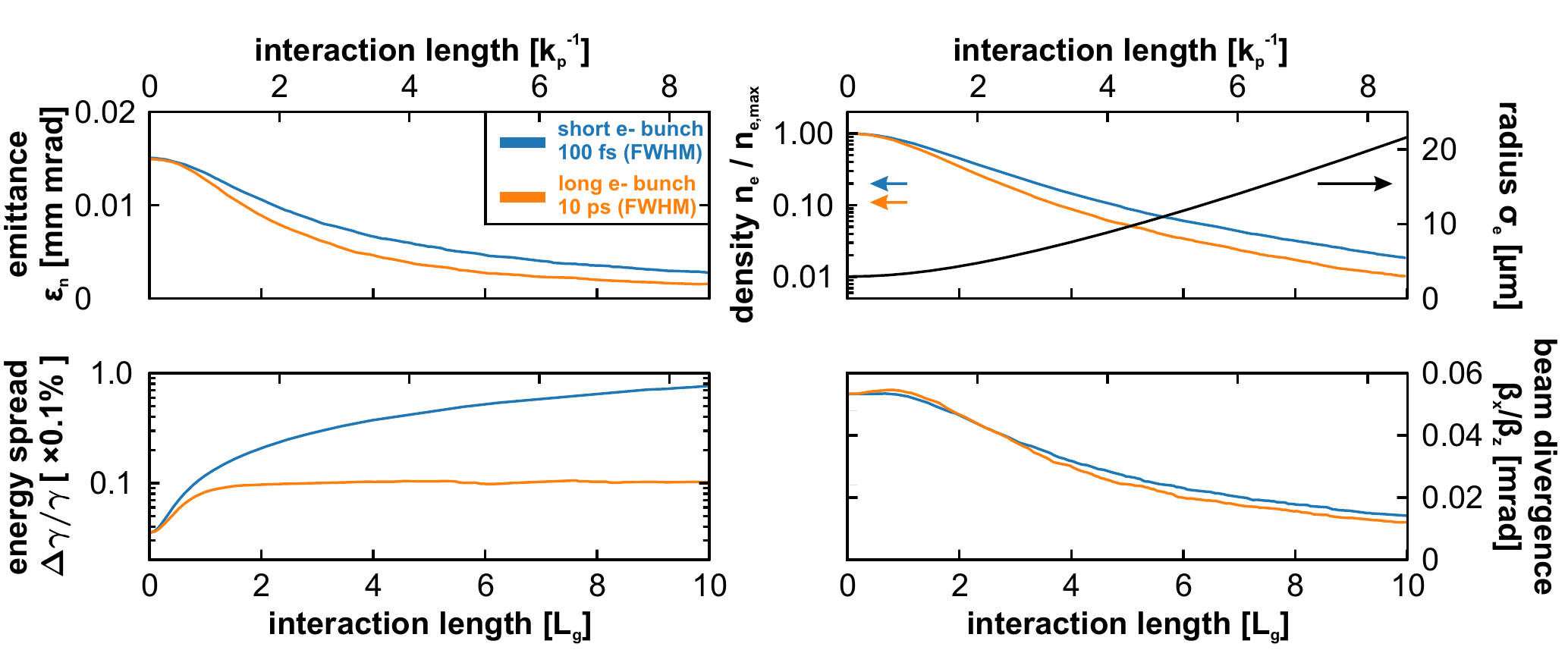}};
    \node[right,align=left] at (0.0,0.0) {(a)};
    \node[right,align=left] at (8.5,0.0) {(c)};
    \node[right,align=left] at (0.0,-3.8) {(b)};
    \node[right,align=left] at (8.5,-3.8) {(d)};
  \end{tikzpicture}
  \caption{Shows the classical beam propagation dynamics including space-charge, but without FEL interaction, for the electron beam of the QFEL example scenario for a short \SI{100}{fs} (orange) beam and a long \SI{10}{ps} beam (blue) simulated. The beam quantities are evaluated within the target region of laser-electron overlap. Any substantial deviation from the initial beam properties destroys QFEL lasing. The beam quantities are (a) transverse emittance, (b) energy spread, (c) beam radius $\sigma_\text{e}$ (rms) common to both scenarios and electron density relative to the initial density and (d) the beam divergence. Except for the beam radius, all these quantities are calculated over a sub-volume of $c\cdot\SI{100}{fs}\times(\SI{5}{\um})^2$ of the beam corresponding to the target QFEL interaction volume defined by the laser beam geometry.}
  \label{fig::QFEL_GPTsims}
\end{figure}

In Fig.~\ref{fig::QFEL_GPTsims} we compare the fully classical beam propagation dynamics of a short (\SI{100}{fs}) and a long (\SI{10}{ps}) electron bunch within the interaction zone without any FEL interaction, subject to space-charge and self-fields only.
The simulations are done with the particle tracker GPT \cite{GPT,VanDerGeer2005} using the \emph{spacecharge3Dmesh} model which calculates space-charge forces between particles based on a classical model.
The laser-electron beam interaction volume for the short bunch is the entire volume, while for the long bunch of same electron density we exclude virtually all longitudinal space-charge dynamics by observing the results in a subset volume corresponding to the short bunch, defined by the laser geometry.
Hence the function of most electrons in the long beam is shielding the core region from longitudinal space-charge forces, but not participating in the QFEL interaction itself.

While there is no scenario in which the electron beam maintains the required electron beam properties in terms of transverse emittance, electron number density and energy spread for even a fraction of the entire QFEL interaction distance, the examples show that low-charge scenarios based on ultra-short electron bunches are dominated by longitudinal space-charge dynamics.
Particularly the relative energy spread in Fig.~\ref{fig::QFEL_GPTsims}(b) deteriorates by orders of magnitudes.

These examples illustrate that for the space-charge dominated beams required for QFEL lasing according to the presented space-charge and spontaneous emission scaling, even ideal electron sources will not suffice to meet this challenge.
In addition to exploiting a (hypothetical) Raman QFEL regime on the microscopic scale, it will be necessary to shield a core part of the interaction region from space-charge forces, either through a sufficiently large and uniform beam or external focusing forces mitigating the space-charge forces.
However, for all practical purposes, such a technology for almost perfectly mitigating space-charge forces along a centimeterm to meter long interaction region currently exceeds the state of the art.

\section{Conclusions}

We conclude that not only any future realization of QFELs faces extraordinary experimental challenges, but also that current theories of QFELs do not account for central aspects relevant for applications.

Central experimental challenges for the electron beam lies in attaining normalized transverse emittances of typically below $0.01-\SI{0.1}{mm.mrad}$ and relative energy spreads below $0.1-\num{1.0e-4}$.
These may be addressed by QFEL designs incorporating low-charge, high-current bunches as suggested in Ref.\ \cite{Bonifacio2017}.

The experimental challenge for the optical undulator lies in providing uniform undulator properties over the entire interaction length.
Intensity variations of the laser pulses providing the optical undulator field typically need to be below a level of $\SI{10}{\percent}$ within the interaction volume.
The choice of a Traveling-Wave Thomson-Scattering (TWTS) geometry can preclude variations originating from the finite Rayleigh length of laser pulses in head-on scattering geometries.
Therefore TWTS allows to efficiently reduce intensity variations below the required level over extended interaction distances.

To this day, there is no theoretical work on Quantum FELs, neither analytical nor numerical, which self-consistently accounts for the full kinetic 3D problem of a QFEL including spontaneous emission and space-charge for realistic, non-ideal electron and laser beams in a rigourous quantum model.
A good modeling of space-charge, transverse coherence, field diffraction, spontaneous emission and 6D electron phase space will be essential to capture the relevant processes affecting the dynamics of a QFEL. Although the derivation of such a highly complex model is a challenge by itself, it will be key to judge on the existence and realizability of Quantum FELs.

The key finding of our study is the exclusion of a deep quantum-mechanical FEL regime in which space-charge forces and spontaneous emission are negligible for the electron dynamics and coherence properties of the QFEL.
This fundamentally excludes the existence of a Quantum SASE FEL without significant space-charge dynamics.
Our argument manifests in the opposed scalings of a minimum and a maximum acceptable optical undulator field strength $a_0$, eqs.\ \eqref{eq::QFEL_Min_a0} and \eqref{eq::QFEL_max_a0} respectively.
The former constraint originates from the requirement of negligible space-charge effects throughout the interaction and the latter by requiring the liftetime of an excited electron momentum state to be comparable to the interaction time.
These constraints reflect that increased gain lengths by modified electron dynamics as well as Coulomb explosion on the electron bunch-scale each prohibit the existence of a Quantum SASE FEL.

We identified two potential loopholes to our argument: First, space-charge blowup could be averted by focusing structures that are effective on a sub-millimeter scale or if a region in the center of an extended, uniform high-charge bunch can be found which is largely unaffected by space-charge.
However, currently such an approach seems to be beyond the state of the art.

Secondly, the electron dynamics including space-charge and spontaneous emission requires a theory of a possible Raman QFEL regime and a complete multi-mode model.
Although first attempts of a description including space-charge were made \cite{Monteiro2013,Serbeto2009,Serbeto2008}, it is not yet clear how gain lengths and bandwidths change or if hard limits to the quantized nature of the FEL regime emerge.
By applying a best-case estimate for the hypothetical Raman QFEL regime, we conclude that the deep quantum regime $\bar{\rho}\leq 0.5$ remains excluded for Quantum SASE FELs.

For potential applications, the central features of a QFEL are to provide a transversally, as well as longitudinally, fully coherent SASE FEL at \si{\angstrom}-wavelength in a relatively compact setup compared to existing x-ray FELs.
Beyond eventual proof-of-concept experiments that mainly explore quantum statistics, Quantum SASE FELs are currently not ready to meet such a demand.

For future theoretical and experimental efforts on Quantum FELs we thus propose to shift the perspective towards seeded Quantum FELs.
One option could be to build a cascade of QFELs each of which operates just over a few gain lengths but starts with a fresh electron bunch to amplify the radiation of the previous QFEL.
In this way the challenge imposed by space-charge effects may be met if the QFEL parameters are tuned for low spontaneous emission.
Analogously to classical FEL theory the saturation length of a seeded QFEL may be expected to become smaller than the one of a Quantum SASE FEL, which will facilitate meeting the challenges brought up in this article.

Hybrid schemes of classical and Quantum FELs are another option.
Especially if one acknowledges the advances of classical FELs, both based on magnetic \cite{Yabashi2017,Pellegrini2016,Barletta2010,McNeil2010,Kim2008} and optical undulators \cite{Steiniger2014_3,Steiniger2014,Steiniger2016,Steiniger2018,Debus2010,Steiniger2014_2} to obtain fully coherent x-ray FELs, as well as compact infrastructure footprints.
In such a scenario Quantum FELs would extend only over a few gain lengths, operate either on a full or filtered coherent (x-ray) seed of a classical FEL source and amplify the incoming seed by several orders of magnitudes.
The benefit of such a QFEL amplifier compared to a classical FEL amplifier is a larger energy conversion efficiency, which allows for a lower electron energy and for more compactness.
Beyond light source applications such seeded Quantum FELs also provide an avenue to meet one of the grand challenges in modern beam diagnostics: the diagnostics of ultra-low emittance beams with high 6D-brightness \cite{Downer2018}, as well as highly coherent x-ray sources.
By exploiting the extreme sensitivity of QFEL physics with regard to non-ideal effects as a diagnostic, Quantum FELs might open a new window for measuring and manipulating the quantum properties of high-brightness electron or x-ray beams.

\section{Acknowledgements}
First of all, we wish to thank Wolfgang Schleich for many years of fruitful cooperation in which we often had the change to benefit from his unique view on physics.
In particular, RS is grateful for a friendship with Wolfgang for more than 30 years while AD and KS appreciate the warm welcome during their visits in Ulm.
We all wish Wolfgang Schleich only the best for his 60th birthday!
We thank M.\ Bussmann, E.\ Giese, R.\ Endrich, P.\ Preiss, A.\ Gover, Y.\ Pan, and P.\ Anisimov for fruitful discussions. 

\newpage
\section{Appendix}

\begin{table}[!htp]
    \centering
    \caption{Overview on Quantum FEL scaling laws and realization criteria.}
    \label{tab::QFELFormulars}
    \scalebox{0.8}{
    \begin{tabular}{lccc}
    \toprule
    FEL characteristics &\\
    \hline
    Optical undulator wavelength & $\lambda_\text{u} = \lambda_0/(1-\beta\cos\phi)$\\
    Normalized optical undulator strength \eqref{eq::a0HandyDef} & $a_0  = \frac{e E_0}{m c \omega_0} \simeq \num{0.8493e-9}\lambda_0[\si{\um}]I_0^{1/2}[\si{W/cm^2}]$\\
    Radiated wavelength \eqref{eq::ThomsonFormular} & $\lambda_\text{FEL}=\frac{\lambda_u (1+a_0^2/2)}{2\gamma_0^2}$\\
    Target electron energy & $\gamma_0=\sqrt{\frac{\lambda_u (1+a_0^2/2)}{2\lambda_\text{FEL}}}$\\

    Classical FEL parameter \eqref{eq::defineRho} & $\rho = \left[ \frac{a_0^2 f_\mathrm{B}^2 \Omega_\text{p}^2}
        {32 \gamma_0^3 c^2 k_0^2 (1-\beta_0\cos\phi)^2}\right]^{1/3}$\\

    QFEL parameter \eqref{eq::QFELparameter} & $\bar{\rho} \equiv \rho \frac{\gamma m c}{\hbar k_\text{FEL}} = \rho\gamma \frac{\lambda_\text{FEL}}{\lambda_c}=\frac{\lambda_c}{\gamma\lambda_\text{FEL}}\sqrt{\frac{\bar{\rho}^3}{1+\bar{\rho}}}$\\

    QFEL gain length \eqref{eq::QFELGainLength} & $L_\text{g}=\frac{\lambda_u}{4\pi\Gamma}$\\
    QFEL saturation length \eqref{eq::QFELSaturationLength} & $L_\text{sat}\simeq 10 L_\text{g} \sim \frac{\lambda_u}{\Gamma}$\\
    QFEL cooperation length \eqref{eq::QFELCooperationLength} & $L_\text{c}=\frac{\lambda_\text{FEL}}{4\pi\Gamma}$\\
    QFEL bandwidth \eqref{eq::gainBandwidthInterpolate} & $\Gamma=\rho\sqrt{\frac{\bar{\rho}}{1+\bar{\rho}}}$\\
    Resulting photon number & $N_\text{e}=Q/e$\\
    Interaction length & $L_\text{int}=N_\text{g}\cdot L_\text{g}$\\
    \hline
    Electron bunch characteristics &\\
    \hline
    Bunch charge & $Q=|e| N_\text{e}$\\
    Bunch duration (rms) & $\tau_\text{e}=Q_\text{e}/I_\text{p}$\\
    Bunch density & $n_\text{e} = I_\text{p} / (2\pi\sigma_\text{e}^2 ec)$\\
    Bunch radius & Choose with $I_\text{p}$ and $\tau_\text{e}$ according to $\rho$ in eq.~\eqref{eq::defineRho}.\\
    & Check against the emittance constraints eqs.~\eqref{eq::emittanceConstraint}, \eqref{eq::defocusingContraint},\\
    & as well as the weaker conditions eqs.~\eqref{eq::defocusingContraint}, \eqref{eq::pondForceConstraint}, \eqref{eq::walkOffConstraint}\\
    \hline
    Quantum FEL constraints &\\
    \hline
    Relative electron bunch energy spread \eqref{eq::energySpreadConstraint}, \eqref{eq::energySpreadConstraint} & $\frac{\Delta\gamma}{\gamma}\leq\Gamma$\\

    Norm. emittance limit for transverse coherence \eqref{eq::QFEL_tranverseCoherence} & $\varepsilon_n = \hat{\varepsilon} \frac{\gamma \lambda_\text{FEL}}{2\pi}$, with $\hat{\varepsilon}\in [0.5 \ldots 10]$\\

    $a_\text{0,min}$ due to space-charge \eqref{eq::QFEL_Min_a0} & $a_\text{0,min} = \sqrt{\frac{8 \lambda_c}{\gamma\lambda_\text{FEL}} (1+\bar{\rho})}$\\
    $a_\text{0,max}$ due to spontaneous emission \eqref{eq::QFEL_max_a0} & $a_\text{0,max} \leq \sqrt{ \frac{3\lambda_c}{\alpha_f\gamma\lambda_\text{FEL}N_\text{g}}\sqrt{\frac{\bar{\rho}^3}{1+\bar{\rho}}}}$\\

    Beam divergence limit \eqref{eq::emittanceConstraint} & $\varepsilon_n\leq\sigma_\text{e}\sqrt{2\Gamma}$\\
    Beam defocusing limit \eqref{eq::defocusingContraint} & $\varepsilon_n \leq \sigma_\text{e}^2\gamma/L_\text{int}$\\

    Irradiance variation limit from wavelength shift \eqref{eq::intensityVariationConstraint} & $\frac{\delta I_0}{I_0}\leq 4\Gamma\frac{1+a_0^2/2}{a_0^2}$\\

    Irradiance limit from ponderomotive deflection \eqref{eq::pondForceConstraint} & $\frac{\delta I_0}{I_0} \le \frac{4}{a_0}\left( \frac{\pi \rho \sigma_\text{e}^2 \gamma^2}{L_\text{int}^2} \right)^{1/4}$\\

    \hline
    Traveling-Wave Thomson Scattering (TWTS) geometry &\\
    \hline
    Emission angle of coherent &  $\phi_\text{sc} \approx \frac{\sin\phi}{2\gamma^2(1-\beta\cos\phi)}(1 + a_0^2/2 )$\\
    radiation for $\lambda_\text{FEL} \ll \lambda_0$\\

    Laser width in interaction plane & $w_y\geq L_\text{int}\sin\phi$\\
    Laser focal width for overlap & $w_x \geq \sqrt{2\pi}\sigma_\text{e}$\\
    Laser pulse duration for overlap & $\tau_0\geq\tau_\text{e} + \frac{\sigma_\text{e}}{c} \tan(\phi/2)$\\
    Laser power ref.~\cite{Steiniger2014_3} & $P_0\,[\text{TW}] = \num{34.29e-3}\text{[TW]} \cdot \frac{a_0^2}{\lambda_0^2}\sigma_\text{e}L_\text{int}\sin\phi$\\

    \hline
    Colliding beam geometry ($\phi=\SI{180}{\degree}$)&\\
    \hline
    Laser spot size & $w_0 \geq \sqrt{2\pi}\sigma_\text{e}$\\
    Interaction length within Rayleigh length \eqref{eq::QFEL_RayleighConstraint} & $L_\text{int} < 2 z_r \text{ with } z_r = \frac{\pi w_0^2}{\lambda_0}$\\
    Maximum laser bandwidth & $(\Delta \lambda_0/\lambda_0)\leq 2\Gamma$\\
    Maximum interaction length due to Gouy phase \eqref{eq::QFELGouyPhaseCriterion} & $\left(\frac{\Delta z}{z_r}\right)_\text{Gouy} = \pm \sqrt{\frac{k_0 z_r 2\Gamma}{1-k_0 z_r 2\Gamma}}$\\
    Maximum usable focal area due to trans. intensity variation \eqref{eq::QFELGaussTransIntensity} & $\left(\frac{\Delta x}{w_0}\right)_\text{Gauss} = \pm \frac{1}{\sqrt{2}}\sqrt{\log{\frac{a_0^2}{a_0^2-4\Gamma-2a_0^2\Gamma}}}$\\
    Minimum laser pulse duration \eqref{eq::QFEL_RayleighConstraint} & $\tau_0=2 L_\text{int}/c$\\
    Maximum usable focal area due to wavefront curvature \eqref{eq::QFELWaveFrontCurvature} & $\left(\frac{\Delta x}{w_0}\right)_\text{curvature} \simeq \pm 4\pi w_0 \sqrt{2\Gamma}/\lambda_0$\\
    Maximum interaction length \eqref{eq::QFELRayleighIntensityCriterion} due to  &\\
    Maximum interaction length due to laser intensity variation within $z_r$ & $\left(\frac{\Delta z}{z_r}\right)_\text{defocus} = \pm \frac{2\sqrt{2\Gamma}}{\sqrt{a_0^2-8\Gamma}}$\\
    \bottomrule
    \end{tabular}
    }
\end{table}%

\bibliographystyle{mybibstyle}
\bibliography{bibliography}

\end{document}